\documentclass[fleqn,usenatbib]{mnras}

\usepackage{newtxtext,newtxmath}

\usepackage[T1]{fontenc}

\DeclareRobustCommand{\VAN}[3]{#2}
\let\VANthebibliography\thebibliography
\def\thebibliography{\DeclareRobustCommand{\VAN}[3]{##3}\VANthebibliography}

\usepackage{graphicx}	
\usepackage{amsmath}	
\usepackage[usenames,dvipsnames]{color}

\newcommand{\kms}{km s$^{-1}$}
\newcommand{\halpha}{H$\alpha$}
\newcommand{\fgas}{$f_\mathrm{gas}$}
\defcitealias{2015ApJ...799..209W}{W15}

\def\Mstar{\hbox{$M_{*}$}}
\def\Mgas{\hbox{$M_\mathrm{gas}$}}
\def\fgas{\hbox{$f_\mathrm{gas}$}}
\def\tdep{\hbox{$t_\mathrm{dep}$}}
\def\Msun{\hbox{M$_{\odot}$}}
\def\sfrunits{\Msun\ yr$^{-1}$}
\def\sfrd{$\Sigma_\mathrm{SFR}$}
\def\Ngal{237}
\def\Ngalgas{63}
\newcommand{\cii}{[C$\scriptstyle\rm~II$]~}     
\newcommand{\ciins}{[C$\scriptstyle\rm~II$]}     
\newcommand{\ci}{[C$\scriptstyle\rm~I$]~}     
\newcommand{\oiii}{[O$\scriptstyle\rm~III$]~}   
\newcommand{\oi}{[O$\scriptstyle\rm~I$]~}   

\title[The evolution of disc dispersion]{Evolution of Gas Velocity Dispersion in Discs from $z\sim8$ to $z\sim0.5$}

\author[E. Wisnioski et al.]{
E. Wisnioski$^{1,2}$,\thanks{E-mail: emily.wisnioski@anu.edu.au}
J. T. Mendel$^{1,2}$, 
R. Leaman$^{3}$, 
T. Tsukui$^{1,2}$, 
H. \"Ubler$^{4,5,6}$,
N. M. F\"orster Schreiber$^{6}$
\\
$^{1}$Research School of Astronomy and Astrophysics, Australian National University, Canberra, ACT 2611, Australia\\
$^{2}$ARC Centre of Excellence for All Sky Astrophysics in 3 Dimensions (ASTRO 3D)\\
$^{3}$Department of Astrophysics, University of Vienna, T\"urkenschanzstrasse 17, 1180 Wien, Austria\\
$^{4}$Kavli Institute for Cosmology, University of Cambridge, Madingley Road, Cambridge CB3 0HA, UK \\
$^{5}$Cavendish Laboratory, University of Cambridge, 19 JJ Thomson Avenue, Cambridge CB3 0HA, UK\\
$^{6}$Max-Planck-Institut für extraterrestrische Physik (MPE), Giessenbachstr. 1, D-85748 Garching, Germany
}

\date{Accepted XXX. Received YYY; in original form ZZZ}

\pubyear{2025}

\begin{document}
\label{firstpage}
\pagerange{\pageref{firstpage}--\pageref{lastpage}}
\maketitle

\begin{abstract}
Together optical/near infrared integral field spectroscopy and resolved sub-millimetre interferometry data have mapped the ionised and molecular gas motions in nearly one thousand galaxies at redshifts $z>0.5$.  While these measurements have revealed a number of key properties about the evolution of disc structure and kinematics, heterogenous techniques and  samples have led to disparate findings - especially when comparing different dynamical tracers (e.g., H$\alpha$, \ciins, CO). In this paper we present a literature compilation of \Ngal~disc galaxies with measurements of velocity dispersion and rotational velocity between $z=0.5-8$, a subset of \Ngalgas~galaxies have measurements of molecular gas fractions.
We explore the connection between disc velocity dispersion measurements over 8 Gyrs as traced by multiple phases with the expectations from Toomre stability models. When sample properties are taken into account (e.g., stellar mass, tracer) there is little evolution in disc dispersions between $z\sim1.5-8$, consistent with expectations from model assumptions. We find ionised gas dispersions are higher by $\sim2\times$ from molecular gas dispersions at a fixed gas mass. These results are sensitive to the molecular gas tracer with results from \cii showing mixed behaviour indicative of its multi-phase origin. The \cii kinematics can be reconciled with molecular and ionised gas tracers when star-formation rates are taken into account.
\end{abstract}




\section{Introduction}
\label{sec.intro}

Early \textit{Hubble Space Telescope} (\textit{HST}) results revealed a high fraction of galaxies with clumpy and irregular morphologies (e.g., \citealt{cowie:1995:10,Glazebrook:1995vn,van-den-bergh:1996:08}). Since these early results, rich multi-wavelength datasets, including imaging, long slit, and integral field spectroscopy, have revealed that the majority of massive galaxies identified at cosmic noon, $z\sim0.5-3$, are rotating discs (e.g., \citealt{2007ApJ...658..763E,2011ApJ...742...96W,2015ApJ...799..209W}). Many discs host transient regions of intense star formation that likely exist for no more than 200-500 Myrs \citep{2014arXiv1410.7398G,2012ApJ...753..114W}. 
In addition to showing structures and rotation consistent with disc galaxies, both the morphological and kinematic data revealed `puffy' discs \citep{2003AA...399..879R,2006ApJ...650..644E,2017ApJ...847...14E} and high line of sight velocity dispersions \citep{2006ApJ...645.1062F,2014ApJ...790...89K,2015ApJ...799..209W} suggestive of thick, turbulent, marginally stable discs. The large scale heights were interpreted as being consistent with gravitational collapse of kpc-sized clumps \citep{2006ApJ...650..644E,2007ApJ...670..237B,2012MNRAS.422.3339W}. 

The \textit{James Webb Space Telescope} (\textit{JWST}) has largely confirmed these morphological results, although the higher resolution imaging of \textit{JWST} reveals that discs were in place at even earlier times \citep{2024ApJ...968L..15K,2023ApJ...942L..42R,2024ApJ...966..113L}.  Comparison with \textit{HST} images reveal that galaxies at $z>1$ are still on average clumpier than at $z=0$ but that the increased resolution and longer wavelengths of \textit{JWST} reveal more regular morphologies \citep{2023ApJ...948L..13J}. The higher resolution allows for more disc features, such as bars, spirals, and lopsidedness, to be explored, revealing a complexity of disc galaxy morphology both in young and old stars \citep{2023arXiv230707599L}. Galaxies are confirmed to already host thick stellar discs at $z\sim4$, when exploring the rest-frame optical/IR light, with typical scale heights of $\sim0.4-0.5$ kpc, albeit with large scatter \citep{2024ApJ...960L..10L, 2024arXiv240915909T}.

Since the first hints of early kinematic discs from ALMA data \citep{2018Natur.553..178S}, there has been a steady growth of results at $z>4$ for cool gas discs (e.g., \citealt{2018Natur.560..613T,2021MNRAS.507.3540J,2021Sci...372.1201T,2021ApJ...911...99F,2022A&A...665L...8H,2023AA...669A..46P}) extended now by \textit{JWST} to also include ionised gas discs \citep{2023ApJ...948L..18N,2023arXiv230207277V,2023arXiv230502478H,2024A&A...687A.115B}. Surprisingly, a number of the observations have revealed, not just a high fraction of galaxies dominated by rotation, but massive early discs \citep{2023ApJ...948L..18N} and dynamically `cold' discs with molecular gas dispersions as low as $\sim15$ km/s \citep{2020Natur.584..201R,2021MNRAS.507.3952R,2021AA...647A.194F}  $-$ seemingly in contention with the ionised gas results at `cosmic noon'. Simulations suggest that the presence of early cold discs could result from co-planer gas accretion \citep{2022MNRAS.510.3266K}, which has been shown to correlate with disc stability \citep{2023MNRAS.524.4346J}. Other simulations suggest that the observational results are consistent with the formation of a thin molecular gas disc where a thicker ionised gas disc forms due to stellar winds and other energy injecta \citep{2019MNRAS.486.1574M, 2022A&A...667A...5R,2024A&A...685A..72K}. Indeed, the multi-phase nature of discs is often ignored in observations due to difficulty in obtaining resolved measurements in different wavebands or across different facilities due to $\sim20$ hour on source integration times. However, some studies compiling samples from the literature have shown a consistent offset between the kinematics of ionised and molecular gas discs \citep{2019ApJ...880...48U, 2021ApJ...909...12G, 2024A&A...689A.273R}.

The evolution of disc galaxies has far-reaching implications with respect to galaxy structure, chemical distribution, and star formation processes. Quantifying the zero-age velocity dispersion, or birth dispersion, of stars can reveal the relative importance of other heating mechanisms over cosmic time (e.g., \citealt{2017MNRAS.472.1879L,2024MNRAS.527.6926M,2023arXiv230304171H}). 
In the context of the Milky Way, many theories and simulations support a `born-hot scenario' in which the early interstellar medium (ISM) of the Milky Way is already turbulent and settles over time as subsequent stellar populations are born \citep{2022MNRAS.514..689B}. More generally, this can be considered as `upside-down' growth \citep{2013ApJ...773...43B,2021MNRAS.503.1815B}. However, other simulations reveal galaxies that form initially as dynamically cold molecular discs in their centers \citep{2022ApJ...928..106T}.  Even if stars are born dynamically `warm,' additional heating is expected to occur through well known internal processes, including GMC scattering \citep{1951ApJ...114..385S} and radial migration \citep{2021MNRAS.507.5882S}. Mergers, e.g., Gaia-Enceladeas in the Milky Way, provide an external heating mechanism that is commonly implemented in simulations (e.g., \citealt{2001ApJ...563L...1F}). However, small variations in mass ratios and growth histories of mergers can have a large effect on final galaxy structures \citep{2023MNRAS.521..995R}. Constraining the relative amount and time of heating mechanisms together with star formation histories has the potential to explain the commonality of structures  across cosmic time (e.g., thin-thick disc dichotomy; \citealt{2019MNRAS.482.3426M}; Leaman et al. \textit{in prep}).
 
Combining the high-redshift studies with other approaches to the cosmic evolution of discs is complicated by observations with limited spatial and spectral resolution. Observations at high redshift are subject to poor spatial resolution relative to the observational beam size or point spread function (PSF). The result, typically referred to as beam smearing, elevates the line of sight velocity dispersion \citep{2011ApJ...741...69D}. The effect is most severe where the velocity gradient, $\Delta V/\Delta R$, is greatest. This occurs at the centre of the galaxy where the star formation peaks, but is dependent on other factors such as inclination, galaxy size relative to beam size, the shape of the PSF, central mass concentration (bulge), etc. \citep{2016ApJ...826..214B}. Forward modelling codes have been developed to account for the beam when fitting disc models (e.g., \citealt{2015AJ....150...92B,2015MNRAS.451.3021D}) however these codes must assume a disc model and often work only on the highest signal to noise data and most regular/symmetric rotators \citep{2016AA...594A..77D, 2024arXiv241107312L}. Forward modelling three dimensional data cannot fundamentally recover intrinsic velocity structures that are poorly resolved due to degeneracies among flux, rotation, and velocity dispersion distribution.

Limits on spectral resolution for optical and near-infrared instruments result in large uncertainties, especially if intrinsic dispersions are below $\sim30$ \kms~(Wisnioski et al. \textit{in prep}). These uncertainties contribute to the larger scatter when looking at population statistics, making it difficult to uncover correlations with key properties (e.g., star formation rates; \citealt{2019ApJ...880...48U}). New instruments are now available (ERIS; \citealt{2018SPIE10702E..09D}) and are being developed (MAVIS; \citealt{2020SPIE11447E..A0E}) with higher spectral resolutions to provide better constraints on dispersions below the resolution limits of past facilities.

In this paper, we aim to unite the mainly optical/near-infrared dispersion results at `cosmic noon' ($0.5<z<3$) with the molecular gas and new \textit{JWST} results out to `cosmic morning' ($3<z<8$; Section~\ref{sec.data}). We compare the data compilation spanning 12 Gyrs with the analytic model in \cite{2015ApJ...799..209W} and provide an updated model using recent literature results (Section~\ref{sec.models}). We discuss the role of gas phase tracer in mapping the evolution of dispersion in Section~\ref{sec.discussion}. We assume a \cite{2003PASP..115..763C} initial mass function and and assume a flat  $\Lambda$CDM cosmology with $\Omega_\mathrm{m}$ = 0.3 and $H_0$ = 70 \kms Mpc$^{-1}$. 

\section{Literature compilation}
\label{sec.data}
In this section we present a heterogeneous data compilation of galaxies observed primarily from $z\sim0.5-8$, corresponding to lookback times of $5-12.8$ Gyrs. We focus on this redshift range due to the apparent tension between recent results. Local galaxies are not included in the compilation. They have been explored in detail in this context (e.g., \citealt{Green:2010fk,2017ApJ...846...35W,2018MNRAS.474.5076J, 2020MNRAS.495.2265V,2021ApJ...909...12G,2022ApJ...928...58L}). For this work, we include galaxies with dispersions measured from resolved spectral features arising from optical emission lines (\halpha, [O$\scriptstyle\rm~III$]) and far-infrared (FIR) / sub-millimetre emission lines (\ciins, \oi, CO transitions). We include measurements that have been made with a variety of instruments and derived with different techniques. We also include lensed and non-lensed galaxies. A brief discussion on the impact of heterogeneous aspect of the data compilation is given in Section~\ref{sec.techniques}. It is worth noting that the datasets for the different gas phases have little overlap and there has yet to be a significant sample of galaxies resolved kinematically with both an ionised and molecular gas tracer beyond $z\sim0$.  In some cases, resolved ionised gas kinematics are available for the same sources as unresolved cold gas measurements (providing gas mass estimates; as discussed in Section~\ref{sub.mmobs}). Table~\ref{tab.samplelist} gives the literature sources used.

\begin{table*}
\caption{Included datasets for this data compilation, $N$ denotes the number of sources, $z$ gives the redshift or redshift range and the last column notes if sources are gravitationally lensed. Some sources have multiple references listed where galaxy properties are taken from multiple sources. Some sources have measurements from multiple lines and are counted twice to reach N=245 total measurments.}
\begin{center}
\begin{tabular}{lrrrrr}
\hline
Paper & $N$ & $z$ & Lensed? & Lines & Measurement\\ 
 &  &  &  &  & Technique$^{a}$\\ 
\hline
optical line tracers\\
\hline
\cite{2013ApJ...768...74T}  &   1 & 1.5     & n & \halpha & data\\
\cite{2019ApJ...880...48U}  & 175 & 0.6-2.7 & n & \halpha & DysmalPy \\
\cite{2024MNRAS.527.9206U}  &   1 & 4.1     & n & \halpha & DysmalPy\\
\cite{2024arXiv240218543F} &   1 & 6.1     & y & \halpha, \oiii & 3D-BAROLO\\ 	
\hline
FIR/sub-mm line tracers\\
\hline%
\cite{2013ApJ...768...74T} &  6 & 1.1-1.5 & n & CO(3-2)& data\\
\cite{2018ApJ...854L..24U} & 1 & 1.4 & n  &CO(3-2) &DysmalPy  \\
\cite{2011ApJ...742...11S} & 1 & 2.3 & y & CO(1-0) & data\\
\cite{2020ApJ...889..141T}, & 1 & 4.3 &	n & \cii & GalPak\\
\cite{2020Natur.581..269N} &  1 & 4.3 & n  & \cii & QubeFit\\ 
\cite{2020Natur.584..201R} &  1 & 4.2 & y  & \cii & Rizzo+18\\
\cite{2021Sci...371..713L} &  1 & 4.8 & n & \cii & 3D-BAROLO\\
\cite{2021Sci...372.1201T} &  1 & 4.4 & n & \cii & data\\
\cite{2021AA...647A.194F}  &  1 & 4.6 & n & \cii & 3D-BAROLO\\	   
\cite{2021MNRAS.507.3952R} &  5 & 4.2-4.7 &	y  & \cii & Rizzo+18 \\
\cite{2021MNRAS.507.3540J} &  6 & 4.4-5.5 & n  & \cii & 3D-BAROLO\\	
\cite{2023AA...669A..46P} &  1 & 6.8 & n  & \cii & 3D-BAROLO \\ 
\cite{2023AA...679A.129R} &  18 & 0.5-3.6 & n  & CO(2-1), CO(3-2), CO(5-4), & 3D-BAROLO\\
 &   &  &  & CO(6-5), \ci & \\
\cite{2023AA...673A.153P} &  {8} & 5.2-7.7 & n  & \cii, \oiii & {KinMS}\\
\cite{2024arXiv240218543F} &  1 & 6.1 & y & \cii & 3D-BAROLO\\ 	
\hline
\end{tabular}
\end{center}
\label{tab.samplelist}
$^\mathrm{a}$ Technique used to measure kinematic parameters: data = data driven techniques including using the outer regions; DysmalPy \citep{2021ApJ...922..143P}; \\
GalPak3D \citep{2015AJ....150...92B}; {KinMS \citep{2013MNRAS.429..534D};} QubeFit \citep{2020ascl.soft05013N}; {Rizzo+18 \citep{2018MNRAS.481.5606R};} 3D-BAROLO \citep{2015MNRAS.451.3021D}.
\end{table*}%

Literature data have been adjusted to a Chabrier IMF for stellar masses. The sample is also heterogeneous in the derivation of these parameters with a different level of constraints due to the availability of photometric bands for spectral energy distribution (SED) fitting. For consistency, we exclude a handful of sources where stellar masses are derived using scaling relations between UV luminosity and stellar mass (e.g., \citealt{2022A&A...668A.121S, 2023AA...673A.153P}). Star formation rates (SFRs) are derived from a number of techniques including optical emission lines, SEDs, and \cii emission. The variety of measurements, which trace star formation of different timescales, may lead to increased scatter (e.g., \citealt{2002A&A...385..454B, 2020A&A...643A...3S}).

We focus on galaxies that have been classified as discs. This focus is for fair comparison in Section~\ref{sec.models} to theoretical models of disc galaxies. We note, however, that this adds bias to the sample. To classify something as a disc is difficult and depends on the adopted definition of `disc' \citep{2017MNRAS.465.1157R,2019ApJ...874...59S}. Differentiating between isolated discs and discs currently undergoing a merger, or resulting from a recent gas-rich interaction, requires either deep imaging and/or high-density spectroscopic surveys currently beyond reach. Distinguishing between a disc and a close pair that is beam-smeared to appear as a disc is typically possible in deep high-quality multi-wavelength data using stellar and gas morphology and information from velocity fields \citep{2008ApJ...682..231S, 2015ApJ...799..209W}. However, many sources, especially at $z\gtrsim3$, don't have deep multi-wavelength data required to be unequivocally classified. It is possible for sources previously identified as discs, or `candidate discs', with kinematics resolved by multiple beams (e.g., \citealt{2022A&A...665L...8H}) to be reclassified as mergers \citep{2024arXiv240719008P} with higher resolution data or information from different wavelengths. For the purposes of this paper and comparison to theoretical models we make a broad definition of `disc' galaxy to include galaxies supported by rotation simply by the measurement of $V/\sigma>1$ or as identified in the original papers. We acknowledge that this likely includes non-virialised discs and discs in the process of merging.

\begin{figure*}
\includegraphics[ scale=0.36,trim=1cm .2cm 0.5cm 0.1cm, clip]{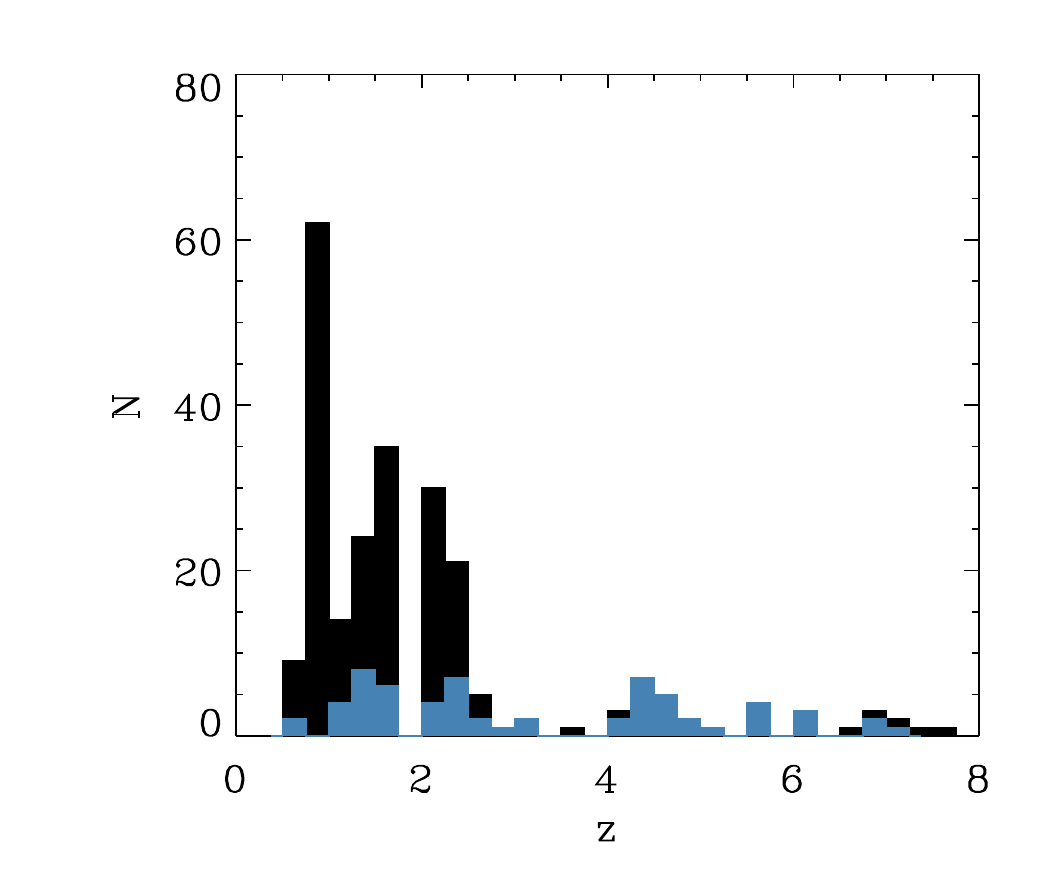}
\includegraphics[ scale=0.36,trim=1cm .2cm 0.5cm 0.1cm, clip]{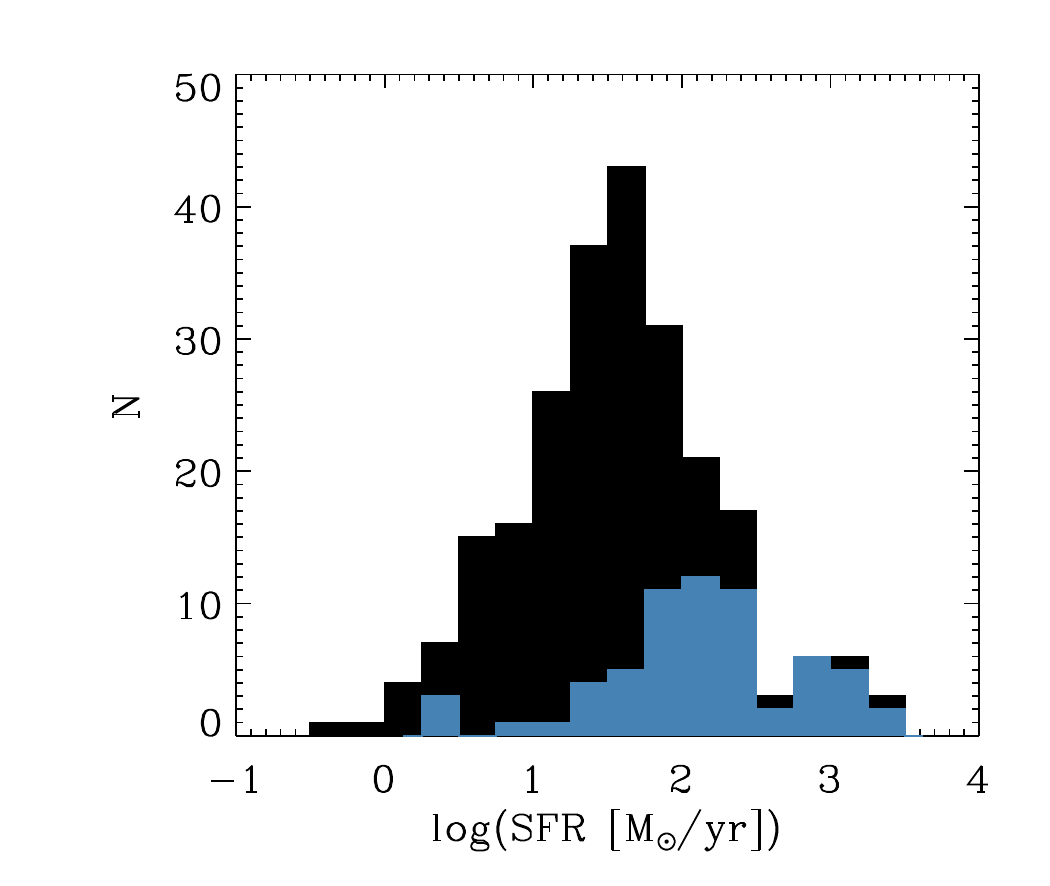}
\includegraphics[ scale=0.36,trim=1cm .2cm 0.5cm 0.1cm, clip]{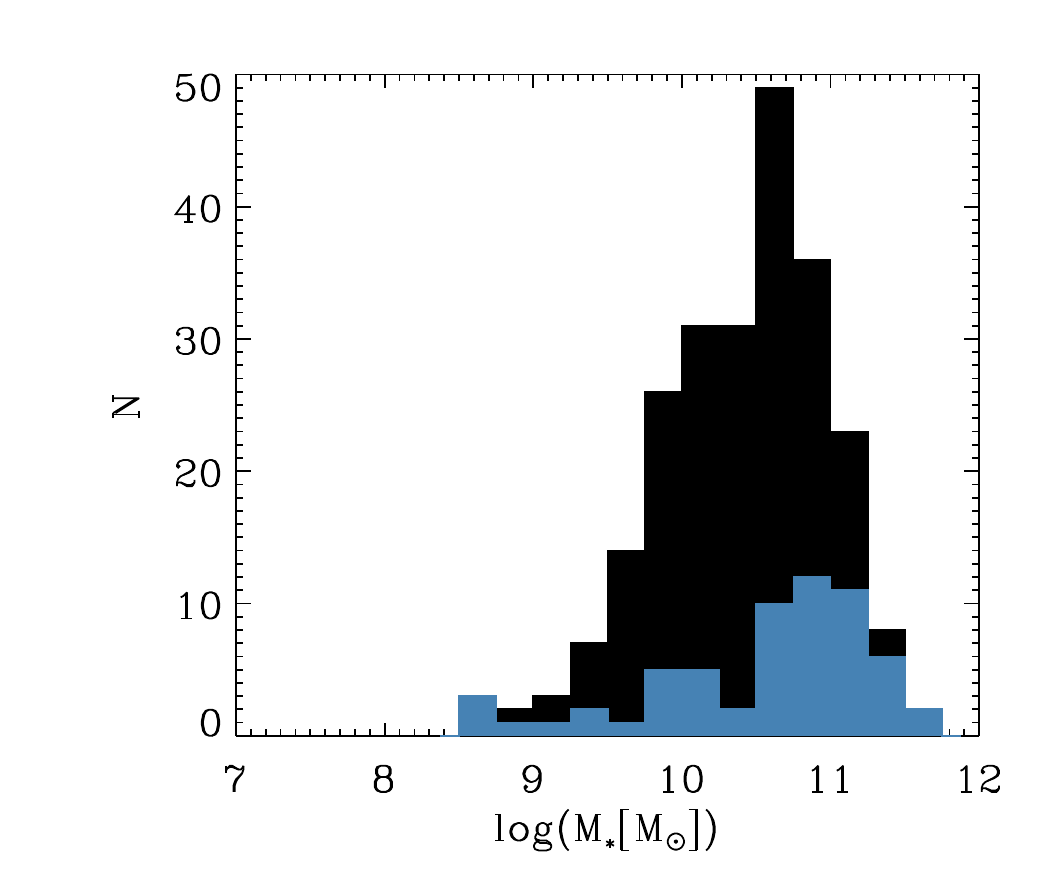}
\includegraphics[ scale=0.36,trim=1cm .2cm 0.5cm 0.5cm, clip]{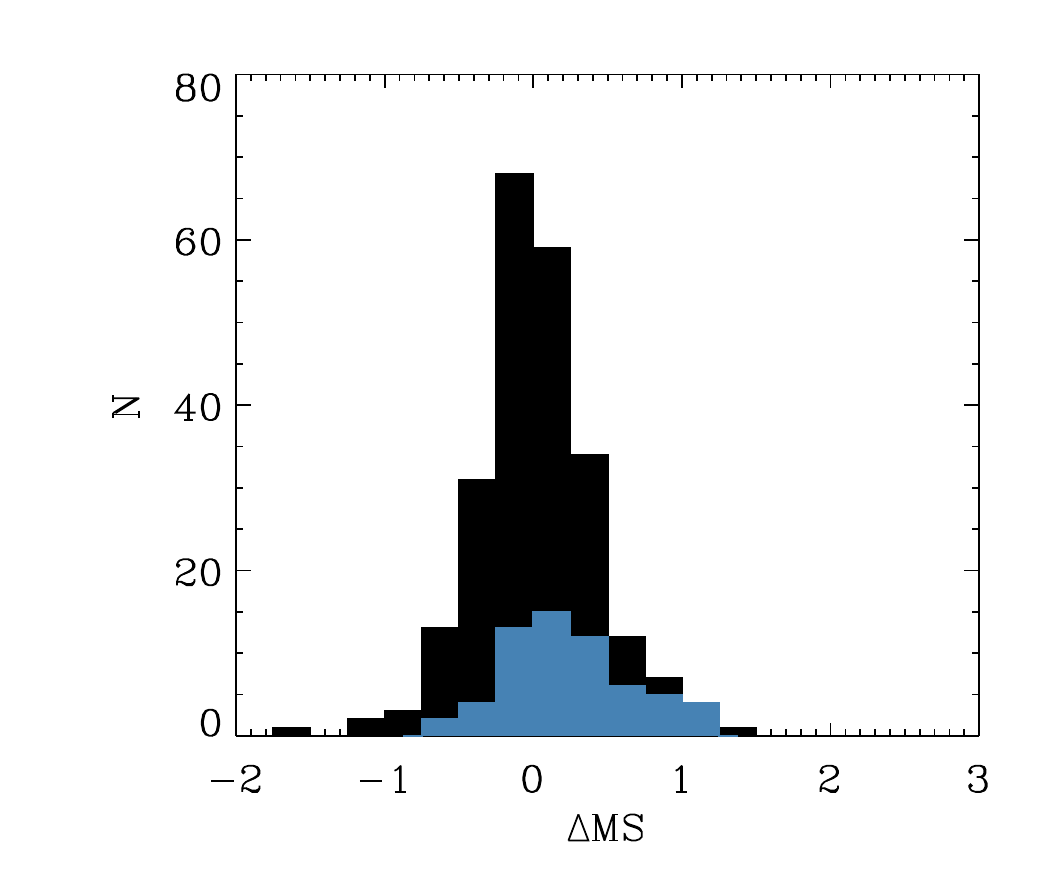}
\includegraphics[ scale=0.36,trim=1cm .2cm 0.5cm 0.5cm, clip]{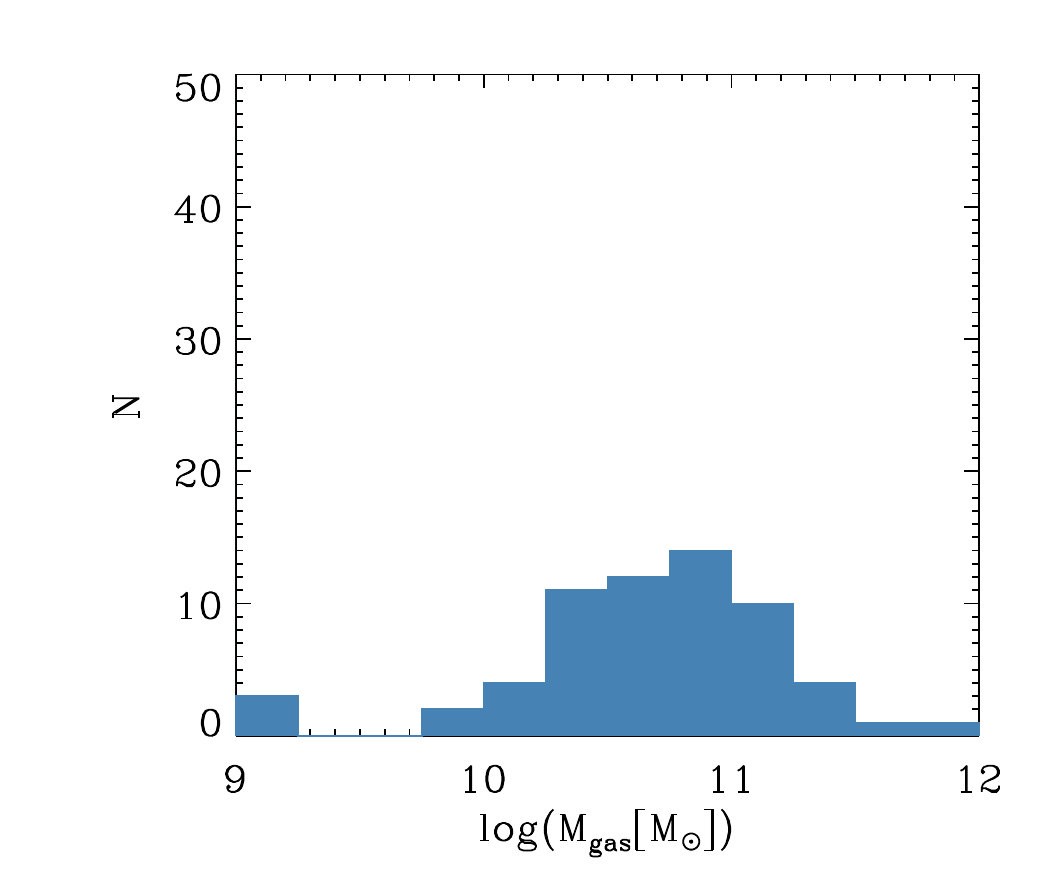}
\includegraphics[ scale=0.36,trim=1cm .2cm 0.5cm 0.5cm, clip]{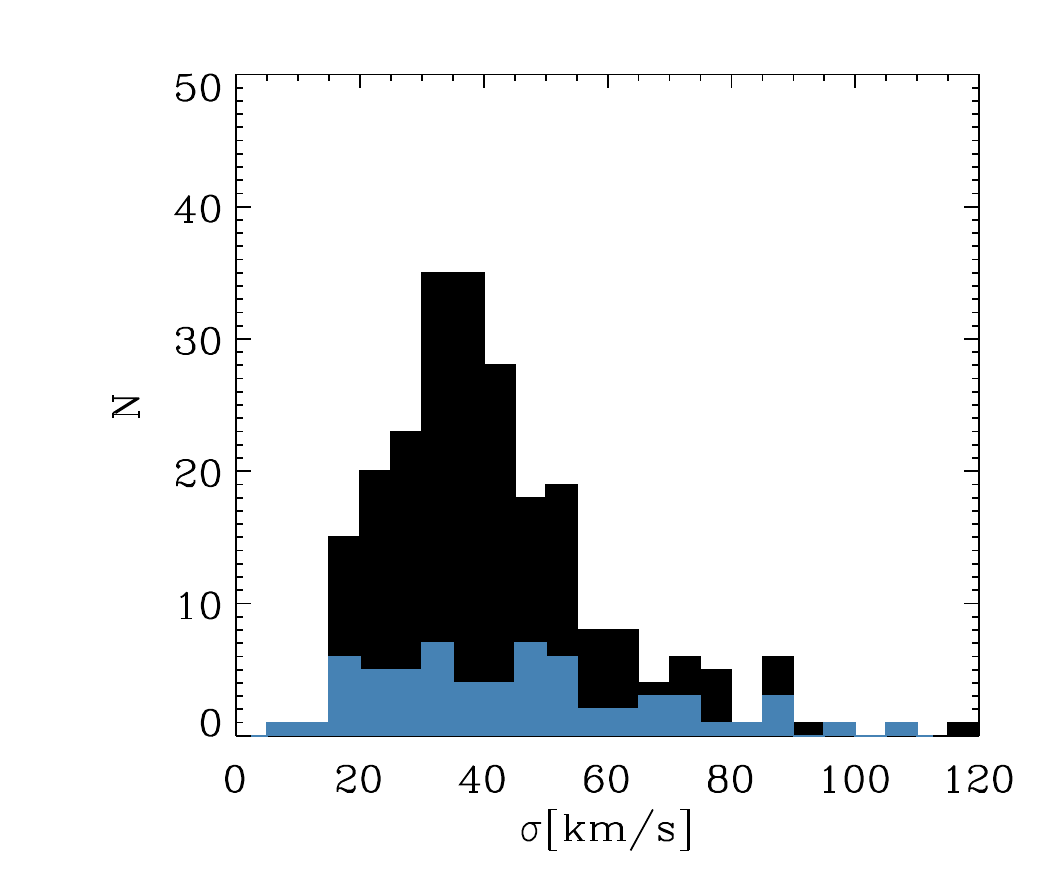}
\caption{Summary of properties of literature compilation. Black histograms indicate the full sample, the blue histograms indicate the sub-sample that have measured molecular gas masses as shown in the middle panel of the bottom row.}
\label{fig.sampleplots}
\end{figure*}

\subsection{Ionised gas with rest-frame optical observations}
Data compilations have been made at cosmic noon by a variety of authors which show good agreement amongst samples \citep{2015ApJ...799..209W,2019ApJ...880...48U}. For this work, we include resolved ionised gas observations at cosmic noon primarily from the KMOS$^\mathrm{3D}$ survey using \halpha. At higher redshifts, we include new JWST results (e.g., \citealt{2024MNRAS.533.4287U}). Although some of these results are not taken with IFUs they are among the first measurements of kinematics of ionised gas discs at $z>4$ available and thus included for an initial comparison with $z>4$ cold gas discs. While the focus of this paper is on kinematics derived from 3D data, we note that agreement is usually seen between high-quality long-slit and IFU data (e.g., \citealt{2014ApJ...790...89K,2016ApJ...819...80P}) with some deviations \citep{2024MNRAS.527.9206U}. Details for the various compiled samples are given in Table~\ref{tab.samplelist}.
Some kinematics, particularly the samples at $z\sim3$ and $z\sim7$ have been derived using forbidden lines, e.g., ~\oiii$\lambda4959,5007$. There has been some work showing that the kinematics of gas traced by the forbidden lines can differ than gas traced by the Balmer lines \citep{2022ApJ...928...58L,2024MNRAS.527.9206U}.

\subsection{Multi-phase gas with millimeter observations}
\label{sub.mmobs}
As with ionised gas, the kinematics of cooler gas has been traced by different emission lines including CO transitions, \ci transitions, and \cii158$\mu$m.  However, these lines are not a direct tracer of the molecular gas and have been shown to trace in some cases a mixture of densities, temperatures, and phases (e.g., \citealt{2018MNRAS.481.1976Z, 2019MNRAS.486.4622C, 2020A&A...643A.141M, 2022MNRAS.517..962D}).  We include the PHIBSS \citep{2013ApJ...768...74T} and ALPAKA \citep{2023AA...679A.129R} surveys and many papers with one or a handful of galaxies. 
The millimeter compilation, primarily coming from the higher redshift universe, $z>3$, is more heterogeneous than the ionised gas compilations as it is difficult to observe large samples due to long integration times. There have also been comparably fewer deep multi-wavelength photometric and spectroscopic surveys to select mass-complete or representative samples.  

The molecular gas masses and stellar masses are also collected for the millimeter data compilation. We have re-derived molecular gas masses where possible using the provided line or dust luminosities to homogenise assumptions and accepted conventions for converting luminosities to molecular gas masses. This allows for consistent assumptions (e.g., $\alpha_\mathrm{CO}$, handling higher J transitions) across datasets. {For galaxies with measurements of resolved [CII] and unresolved CO we utilise the [CII] measurements for the kinematics but preference any unresolved CO measurements for estimates of molecular gas masses.}

For galaxies with available CO measurements, we started with flux measurements reported in the literature. Converting from high-J CO flux to CO(1-0) flux, we adopt Table 2 of \citet{carilli:2013aa} for quasars (QSOs) and sub-milimetre galaxies (SMGs; as identified in the original papers). The compilation includes 16 SMGs and 2 QSOs. For normal main sequence (MS) galaxies, we used $R_{21}=0.9$ \citep{carilli:2013aa}, $R_{31}=0.5$ adopted by \cite{2013ApJ...768...74T}, $R_{41}=0.25$ from the $R_{31}$ value and the measured ratio $R_{43}=0.54\pm0.15$ for DYNAMO $z=1-2$ MS analogue sample \citep{2023ApJ...945....9L}. We then convert the CO(1-0) flux luminosity to total molecular gas mass with $\alpha_\mathrm{CO}\sim0.8$ for both SMG and QSO and $\alpha_\mathrm{CO}\sim4.36$ for normal MS star-forming galaxies. We assumed $40\%$ uncertainty on high-J to CO(1-0) conversion, $40\%$ on $\alpha_\mathrm{CO}$, and statistical uncertainty of the flux, finding the total uncertainty on the molecular gas mass in root-sum squared. The mean uncertainty derived is $60\%$, consistent with the assumed systematic uncertainty of $50\%$ used in \cite{2013ApJ...768...74T}. 

The value of $\alpha_\mathrm{CO}$ is inferred to be consistent with the Milky Way value for normal star-forming galaxies and much less for ULIRGs, SMGs, and QSOs based on dynamical arguments \citep{2008ApJ...680..246T,2012ApJ...760...11H}. The works calibrating the conversion factor including dust measurements also suggest similar trends, and relatively small values for galaxies above the main sequence \citep{2015ApJ...800...20G}. 

For galaxies with available \ci($1-0$) and \cii flux measurements but without CO measurements, we assume that these line fluxes are molecular gas mass tracers, with the conversion factors $\alpha_\mathrm{CI}=17$ with $20\%$ uncertainty based on calibration by \cite{2022MNRAS.517..962D}, and conversion factor by \cite{2018MNRAS.481.1976Z}, $M_\mathrm{mol}/L_\mathrm{[CII]}=30$ with $50\%$ uncertainty. For the gravitationally lensed systems, we used the reported {intrinsic luminosity after correcting the lensing magnifications} \citep{2021MNRAS.507.3952R,2024arXiv240218543F}.

There can be large uncertainties associated with molecular mass measurements resulting from unknown metallicity and dust dependencies (see, e.g., \citealt{2023arXiv230307376E}). In this study, we do not include any possible variation in the conversion factor due to metallicity which is unknown for the majority of the sample. 

We also cross-match the ionised gas data compilation at cosmic noon with published molecular gas or dust measurements from \cite{2019ApJS..244...40L}, \cite{2020ApJ...901...74T}, \cite{2020ApJ...899...37K}, and \cite{2024A&A...685A...1A}. A total of 11 matches were identified and homogenised as described above.

The distribution of the total sample, including the optical, FIR, and millimeter literature compilations, is shown by by black histograms in Fig.~\ref{fig.sampleplots}. Parameters include redshift, star formation rate, stellar mass, offset from the main sequence, molecular gas mass, and disc velocity dispersion. The offset from the main sequence ($\Delta$MS) is measured with respect to the main sequence defined by \cite{2014ApJS..214...15S} for consistency with Section~\ref{sub.ssfr}. The full sample is representative of the massive galaxy population at these redshifts (e.g., median $\Delta$MS $=-0.09$, median $\log$[\Mstar] $=10.4$). The subset of galaxies with estimated molecular gas masses, 26\%, are over-plotted in blue. This subset spans the full range of redshift and velocity dispersions but is concentrated at higher star formation rates, higher stellar masses, and to on or above main sequence galaxies (median $\Delta$MS$ =0.34$) reflecting the demanding molecular gas line observations for low mass systems.

\subsection{Kinematic measurement techniques}
\label{sec.techniques}
Kinematic properties, specifically disc velocity dispersion ($\sigma$) and rotational velocities ($V$), have been measured differently across the literature. While some works measure values from 1D or 2D fits and apply a beam smearing correction, others use forward modelling techniques applied to the 3D data cubes. Beam smearing corrections and modelling codes rely on underlying assumptions of the kinematics (e.g. rotation model, radial variation in dispersion) that also vary across the literature.  For the purposes of this work we have not attempted to homogenise methods. This undoubtedly results in a higher scatter in the combined dataset (e.g., \citealt{2011ApJ...741...69D}). It is beyond the scope of this paper to re-fit all kinematic results in the literature at the cube level with the same kinematic tool and or methods. Future work will employ a non-parametric model to available public IFU data (Kanowski et al. \textit{subm}).

Some investigations have shown comparisons between measurement techniques (e.g., \citealt{2011ApJ...741...69D,2019MNRAS.485.4024V,2023AA...673A.153P, 2024arXiv241107312L}) however the comparisons are subject to model input and data quality, including S/N, spatial resolution, spectral resolution, error handling, etc.. We show the variety of measurement techniques and emission lines of the full data compilation in the next Section. 

Finally, we note that while most codes seem to agree when extracting observed velocities, inclination uncertainties can have a large impact on derived rotational or circular velocities (where rotational velocity is the observed velocity corrected for inclination and beam-smearing and circular velocity also includes a pressure-support term). Inclinations are typically derived from the highest resolution imaging available, which can range from \textit{JWST} to the kinematic data itself. For galaxies with only a few resolution elements, it is difficult to accurately constrain the structural morphology, especially if galaxies are thicker or more triaxial at early times (e.g., \citealt{2014ApJ...792L...6V, 2017ApJ...847...14E,2023arXiv230304171H}). 

\section{Results and Analysis}
\label{sec.models}

We plot the dispersion measurements as a function of redshift for the full literature compilation in Fig.~\ref{fig.Tkeyplot}. Each panel shows the compilation color-coded by different observational or technical characteristics to highlight the heterogeneous nature of the sample. The same figure is reproduced in Fig~\ref{fig.age} as a function of lookback time, rather than redshift, and with dispersion visualised on a linear scale.  There is a clear split with redshift with regards to which gas tracer is predominately measured. Observations with \textit{JWST} will reduce this bias for ionised gas observations at $z>4$, however with current mm/sub-mm facilities the the same line can not be used to trace molecular gas across the full redshift range. While we do not see an obvious bias introduced by the different codes used in this heterogeneous dataset, Fig.~\ref{fig.Tkeyplot} (bottom), some works have made direct comparisons and found systematic offsets, e.g., \cite{2023AA...673A.153P,2024arXiv241107312L}. We denote measurements from sources that are gravitationally lensed with a star in the bottom panel.

The results show considerable scatter around a median of 40 \kms, with little evolution between $z=1$ to $z=8$. The flat evolution of $\sigma_\mathrm{gas}$ has been seen at high redshift from other authors \citep{2023AA...679A.129R, 2024A&A...689A.273R}. However, it is in contrast to expectations of what would have been extrapolated from previous model assumptions (red hatched) extending to $\sim100$ \kms~at $z\sim6$ (e.g., \citealt{2023AA...672A.106L,2023AA...669A..46P}). This model, proposed in \cite{2015ApJ...799..209W}, hereafter, \citetalias{2015ApJ...799..209W}, built on Toomre \citep{1964ApJ...139.1217T} stability theory for disc galaxies. In this section we explore the model assumptions used in \citetalias{2015ApJ...799..209W} and provide updated prescriptions for variables that evolve as a function of time. The updated model, described below, is shown by the gray band in Fig.~\ref{fig.Tkeyplot} which shows better agreement with the full data compilation.

\subsection{Toomre stability model}
\label{sec.toomre}
An assumption invoked to reconcile observations of high gas velocity dispersions in galaxies at $z>1$ with theoretical expectations is that the gas is in a state of marginal gravitational stability parameterised by the Toomre parameter, $Q$, where
\begin{equation}
Q_\mathrm{gas}=\frac{\sigma_\mathrm{gas}\kappa}{\pi G\Sigma_\mathrm{gas}}\approx1
\label{eq.toomre}
\end{equation}
for a single phase of gas. Above, $\Sigma_\mathrm{gas}$ is the surface density, $\kappa$ is the epicyclic frequency, $G$ the gravitational constant, and $\sigma_\mathrm{gas}$ is the radial velocity dispersion of the gas.
This simplified argument has been used to explain the evolution of gas velocity dispersion assuming isotropic (or radially constant) gas velocity dispersion (e.g., \citealt{2011ApJ...733..101G,2015ApJ...799..209W}) and the existence of large star-forming clumps at early times (e.g., \citealt{2004ApJ...611...20I,2011ApJ...733..101G,2012MNRAS.422.3339W}). 
To directly compare with kinematic results, \cite{2011ApJ...733..101G} derived the Toomre relation in the form of 
\begin{equation}
Q_\mathrm{gas}=\frac{\sigma}{V}\frac{a}{f_\mathrm{gas}}
\label{eq.toomre2}
\end{equation}
where $a$ describes the rotation model with a values of 1, $\sqrt{2}$, $\sqrt{3}$, and 2 for a Keplerian, constant rotation velocity, uniform density, and solid body disk, and $V$ is the rotational velocity, where $V=V_\mathrm{obs}$/sin($i$) and $i$ is the inclination. For simplicity, we do not make a pressure support correction (e.g., \citealt{2010ApJ...725.2324B}).
To derive the mass-average evolution of dispersion over time $t$, $\sigma(t)$, \fgas$(t)$ can be parametrised as
\begin{equation}
\fgas(t) = \frac{1}{1+(\tdep(t) \mathrm{sSFR}(t))^{-1}},
\label{eq.fgas}
\end{equation}
where sSFR$(t)$ is the specific star formation rates, \tdep($t$) is the depletion time, and $f_\mathrm{gas}(t)$ is the gas fraction.  
 
While this theory successfully reproduces a number of observations it is a simplification in many respects which are explored in the following sections. The assumption of $Q\approx1$ in particular, and treatment of only a single phase of gas is discussed in Section~\ref{sub.other}. { The use of this derivation is to relate the changing conditions of galaxies over cosmic time, e.g. more molecular gas and higher star-formation rates, to the apparent change in disc velocity dispersions over time. Given the heterogeneous nature of the data compilation in both quality and phase we do not explore more complex derivations of Toomre stability theory here (e.g. \citealt{1994ApJ...427..759W, 2010MNRAS.407.1223R, 2011MNRAS.416.1191R, 2023MNRAS.518.5154N, 2023MNRAS.522.2543A,2024MNRAS.532.3839A, 2024A&A...687A.115B} ). A model exploration will be published in a future work (Leaman et al. in prep). }

\begin{figure}
\includegraphics[ scale=0.5,trim=1cm 3.0cm 0.0cm 0cm, clip]{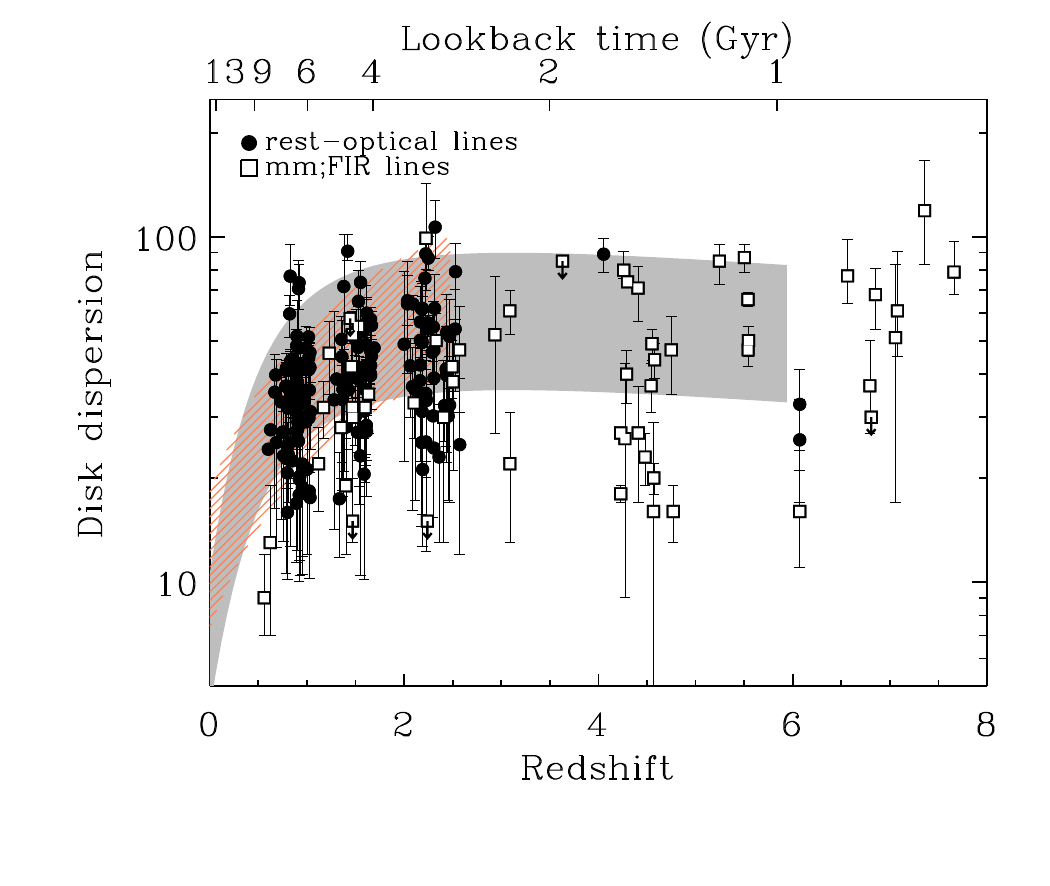}
\includegraphics[ scale=0.5,trim=1cm 3.0cm 0.0cm 1.5cm, clip]{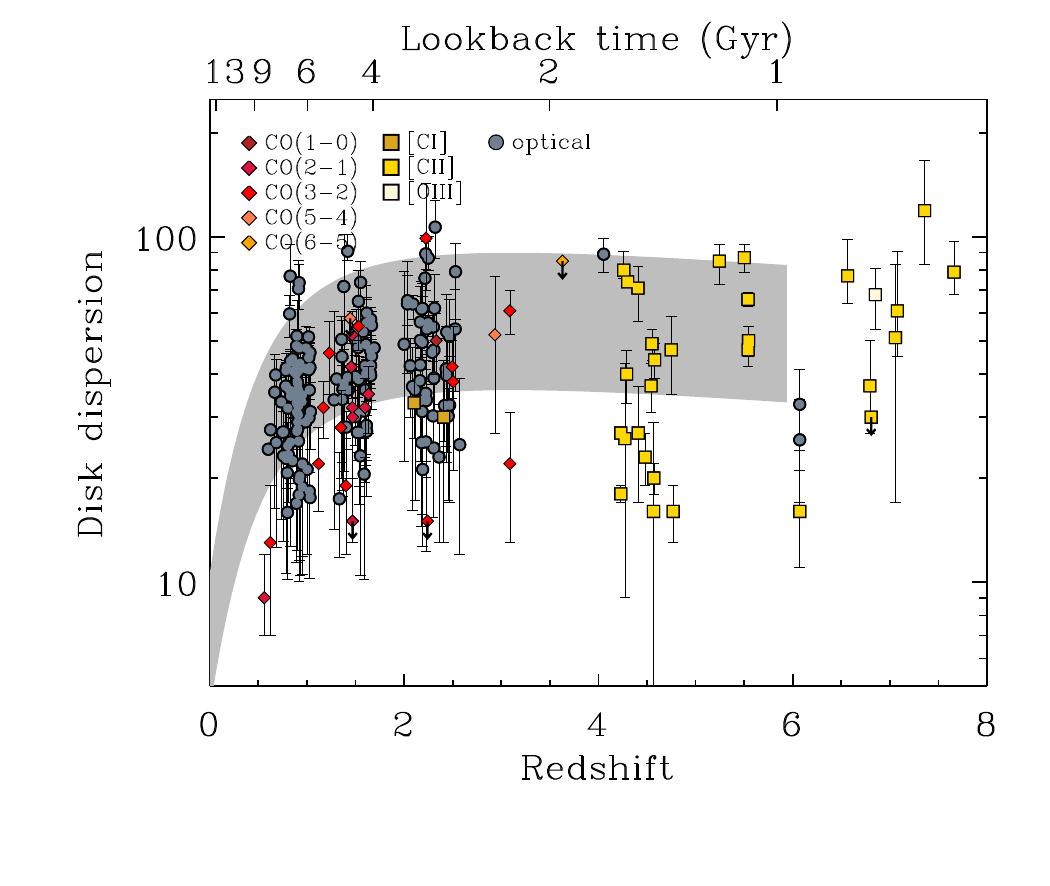}
\includegraphics[ scale=0.5,trim=1cm 1.4cm 0.0cm 1.5cm, clip]{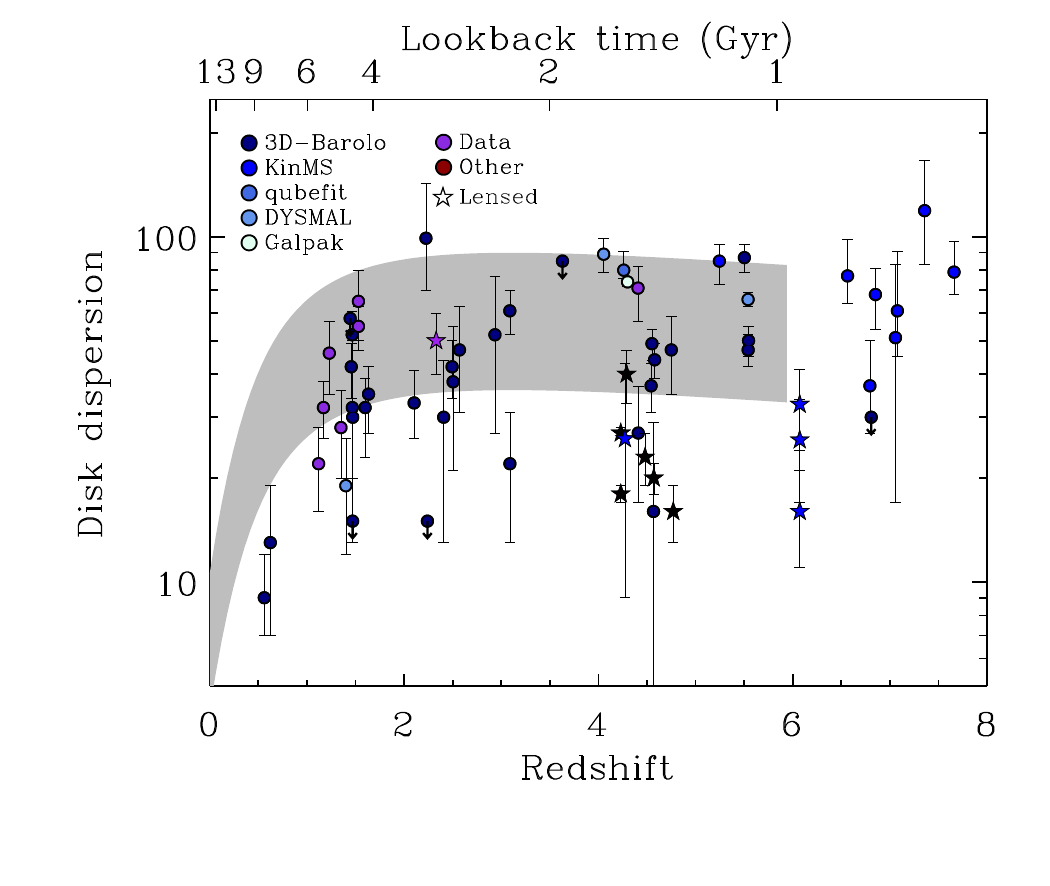}
\caption{Ionised and molecular gas data compilation at $z>0.5$ of disc velocity dispersion. \textit{Top:} The black circles indicate measurements from ionised gas with primarily integral field spectroscopic data. The white squares indicate measurements from resolved molecular gas interferometric data. Upper limits in both cases are indicated with downward arrows. The gray and red bands show predictions from a simplified Toomre stability model. The gray band represents an update from \citetalias{2015ApJ...799..209W} (red dashed) using more recent data-driven prescriptions for sSFR(z) and \tdep(z). The bands are shown only at the redshifts which sSFR(z) and \tdep(z) have been reliably measured. \textit{Middle:} Same as the top panel with symbols coded by emission line. The \ci and \cii group includes \ci(1-0) 609 $\mu$m, \ci(2-1) 370 $\mu$m, and \cii158 $\mu$m lines. The \oiii group refers to the \oiii88 $\mu$m FIR line. \textit{Bottom:} Same as above, for mm/FIR sample, with symbols coded by kinematic extraction technique. Lensed galaxies are shown by stars while all other data are shown by circles.  Purple points show measurements extracted directly from the data without using 3D modelling codes. 
We include data analysed with 3D modelling codes including, 3D-Barolo \protect\citep{2015MNRAS.451.3021D}, KinMS \protect\citep{2013MNRAS.429..534D}, qubefit, DYSMAL, and GalPak3D \protect\citep{2015AJ....150...92B}.}
\label{fig.Tkeyplot}
\end{figure}

The main difference between the updated model (gray) and model from \citetalias{2015ApJ...799..209W} (red) in Fig.~\ref{fig.Tkeyplot}~comes from the assumptions for the evolution of sSFR$(t)$ and $f_\mathrm{gas}(t)$ or  \tdep($t$). Each of these has a secondary dependence on stellar mass.  In \citetalias{2015ApJ...799..209W}, sSFR$(t)$ and $f_\mathrm{gas}(t)$ were parametrised using \cite{2014ApJ...795..104W} and \cite{2013ApJ...768...74T} to $z\sim2.5-3$, respectively. 
For the main body of this paper we adopt \cite{2020ARA&A..58..157T} for $t_\mathrm{dep}(t)$ and \cite{2014ApJS..214...15S} for sSFR$(t)$ (as used by \citealt{2020ARA&A..58..157T}). We explore the different parameterisations of sSFR($t$,\Mstar) and \tdep($t$,\Mstar) in Sections~\ref{sub.molgas} and \ref{sub.ssfr}.

\begin{figure*}
\begin{center}
\includegraphics[ scale=0.73,angle=90,trim=1.05cm 0cm 0.2cm 0cm, clip]{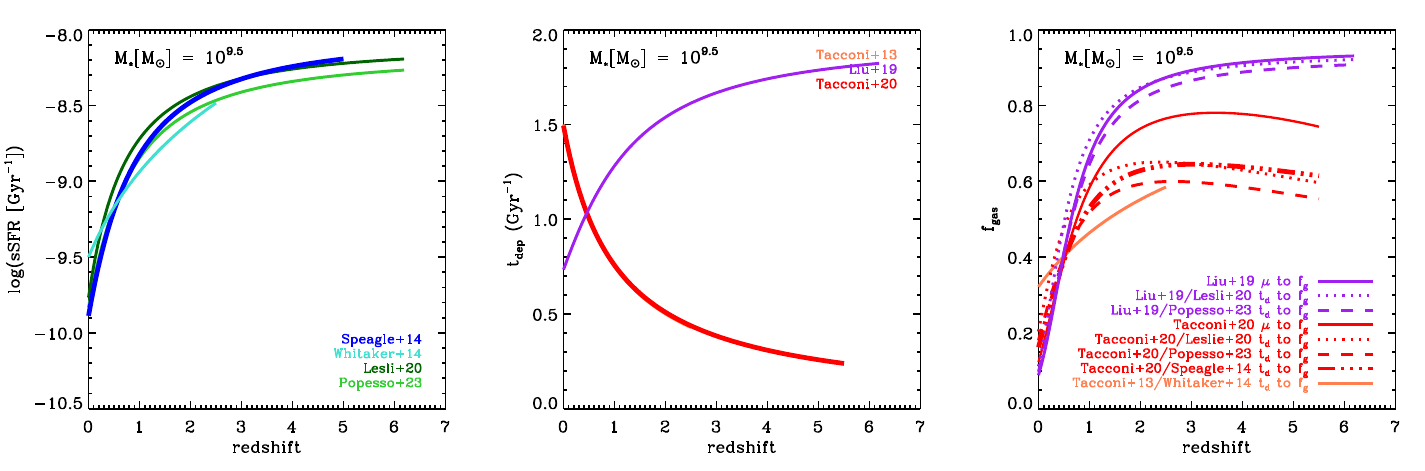}
\includegraphics[ scale=0.73,angle=90,trim=1.05cm 0cm 0.2cm 0cm, clip]{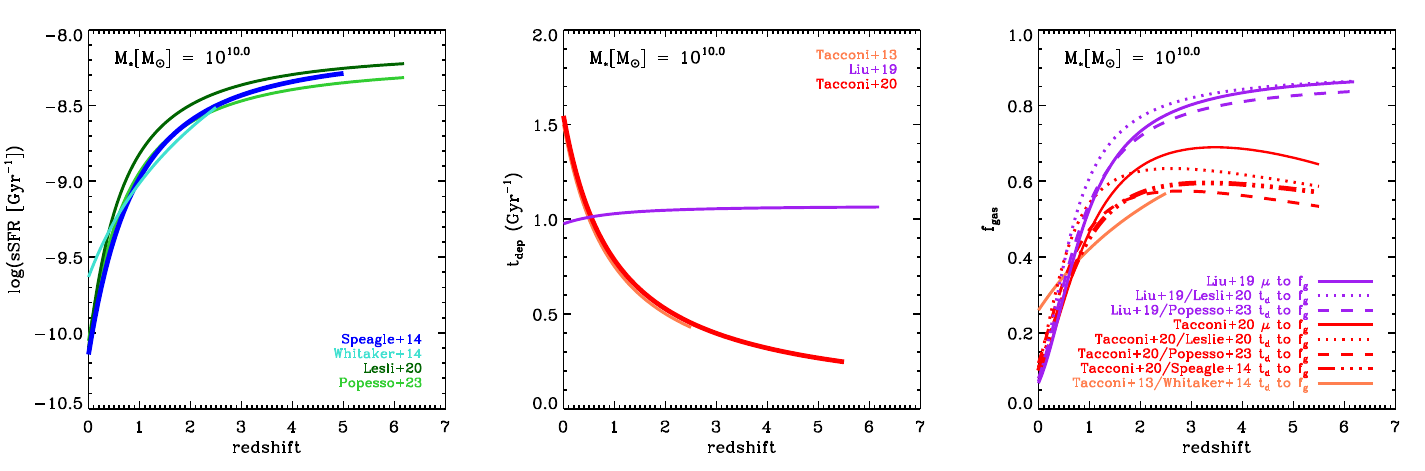}
\includegraphics[ scale=0.73,angle=90,trim=1.05cm 0cm 0.2cm 0cm, clip]{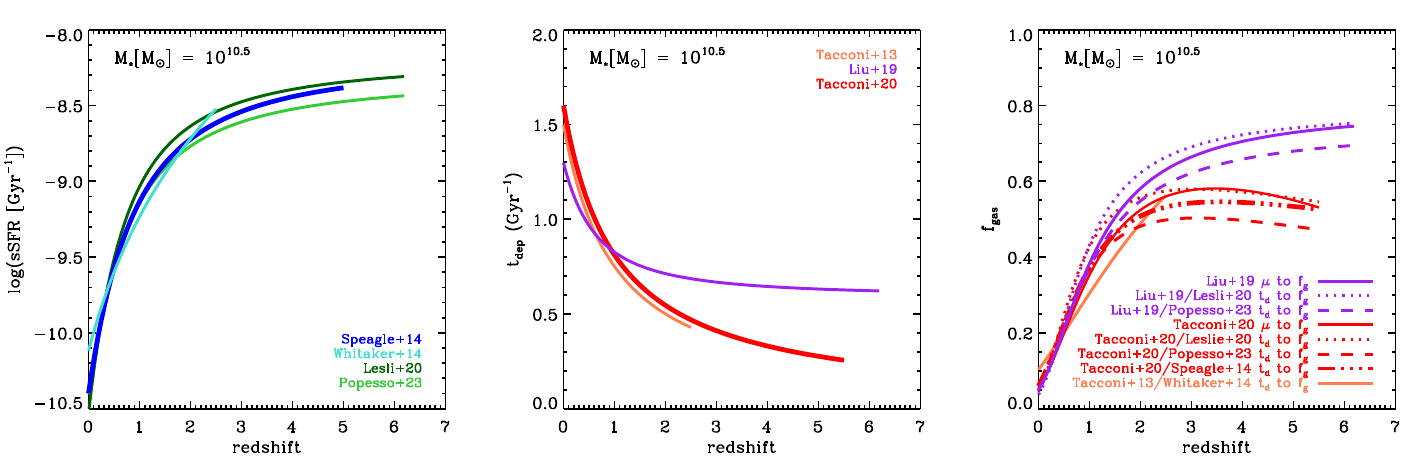}
\includegraphics[ scale=0.73,angle=90,trim=0.3cm 0cm 0.2cm 0cm, clip]{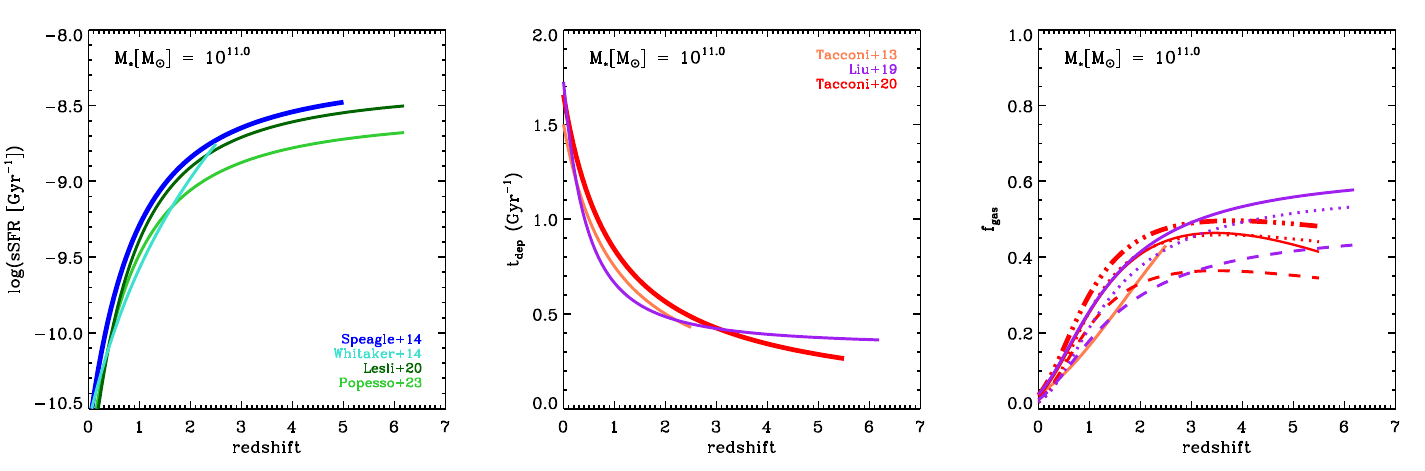}
\caption{ A sub-sample of commonly adopted evolutions of sSFR($z$) (left), $t_\mathrm{dep}$($z$) (middle), and $f_\mathrm{gas}$ (right) at a stellar mass of log($M_*$[\Msun])=[10.0,10.5,11.0] from top to bottom respectively. Lines show the extent of the datasets used. For sSFR(z) (left) we include comparisons of \protect\cite{2014ApJS..214...15S}, \protect\cite{2014ApJ...795..104W}, \protect\cite{2020ApJ...899...58L}, and \protect\cite{2023MNRAS.519.1526P}. Relations are extended to the maximum redshift of the data included in the respective samples. In the middle column we compare $t_\mathrm{dep}$($z$) prescriptions from \protect\cite{2013ApJ...768...74T}, \protect\cite{2019ApJ...887..235L}, and \protect\cite{2020ARA&A..58..157T}. We note that the different references adopt different SFR indicators which can also lead to differences in depletion times. In the right column we compare different derivations for the evolution of gas fractions. As described in Section~\ref{sub.molgas}, $f_\mathrm{gas}$($z$) can be derived using equation~\ref{eq.toomre} or converting $\mu$($z$), where $\mu$ equals $M_\mathrm{gas}/M_\mathrm{*}$, to $f_\mathrm{gas} =  M_\mathrm{gas}/(M_\mathrm{*}+M_\mathrm{gas})$. We assume the SFR/SFR(MS) = 1 for all included derivations for simplicity but note that the more recent derivations of $t_\mathrm{dep}$ and $f_\mathrm{gas}$ do include a dependency on offset from the main sequence. We do not include all possible literature derivations of these properties but pick a relevant subset to show the magnitude of differing assumptions. We adopt \protect\cite{2014ApJS..214...15S} for sSFR($t$) and \protect\cite{2020ARA&A..58..157T} for \tdep in the remainder of the paper and highlight these relations in a bold line.}
\label{afig.comp}
\end{center}
\end{figure*}

\subsection{The role of molecular gas \& depletion time evolution}
\label{sub.molgas}

Large compilations of molecular gas estimates from the literature using emission lines from CO transitions and FIR fine structure lines as well as dust continuum measurements have been use to map the evolution of molecular gas properties across time, mass, and star-formation rates (e.g., \citealt{2017ApJ...837..150S,2020ARA&A..58..157T,2019ApJ...887..235L}).  A detailed comparison of molecular gas evolution is given in \cite{2019ApJ...887..235L} and \cite{2020A&A...643A...5D}.

In short, differences in the evolution prescriptions can result from sample selection (e.g., $z$, $M_*$, SFR) and molecular gas tracers. The functions defined by \cite{2018ApJ...853..179T} and \cite{2017ApJ...837..150S} are primarily derived using data from $z=0-3$ and should therefore not be extrapolated, while \cite{2019ApJ...887..235L} and \cite{2020ARA&A..58..157T} extend to $z\sim4.5$. Recent work by \cite{2020A&A...643A...5D} at $z\sim5-6$ with the ALPINE-ALMA survey of moderate mass ($M_*\sim10^{9}-10^{10}$ \Msun) galaxies using \ciins, supports the extension of the \cite{2018ApJ...853..179T,2020ARA&A..58..157T} results to $z\sim6$. The evolution of gas fractions and depletion times do show some dependence on stellar mass which is seen in all the above works but is particularly pronounced in \cite{2019ApJ...887..235L}. In \cite{2019ApJ...887..235L} the depletion time shows a reversal in slope towards lower masses. The derived evolution of depletion time and molecular gas fraction for different stellar mass bins are shown in the middle and right panels of Fig.~\ref{afig.comp} respectively.

Given the evidence that the equations in  \cite{2020ARA&A..58..157T} can be extended to $z\sim6$ we adopt the depletion time scaling relation of equation 4 from \cite{2020ARA&A..58..157T} which characterises depletion time as a function of redshift, stellar mass, and MS offset. The inclusion of offset from the star-formation main sequence ($\Delta$MS) in \tdep($t$,\Mstar) can change the normalisation of $t_\mathrm{dep}(z)$ but not the slope. Taking into account the MS term removes the need for a varying $\alpha_\mathrm{CO}$ for $\Delta$MS. For simplicity we ignore this term assuming all galaxies are `main sequence' galaxies. Most galaxies in the literature sample fulfil this criteria within errors as shown in Fig.~\ref{fig.sampleplots}. However the distribution has outliers and extends beyond 1 dex of the MS.  
Offsets from the MS could play a large role in terms of the amount of molecular gas present (e.g., \citealt{2020ARA&A..58..157T}). For example, using $\Delta$MS = [-0.5, 0.5] at $\log$\Mstar[\Msun]=10.5 would result in a factor of [0.5,1.4] in \fgas~and $\sigma$. We explore some of our key results in Appendix~\ref{app.dMS} with respect to offsets from the MS.

We note that in this section we explore predictions for velocity dispersion of galaxies at fixed mass at different redshifts but the evolutionary pathways for \textit{individual} galaxies is mass dependent. Today's most massive galaxies (in more massive halos; $M_\mathrm{halo}=10^{14}$ at $z=0$) likely had a relatively flat molecular gas fraction until $z\sim2$ followed by a decline. In contrast, the evolutionary pathways of less massive galaxies (in $M_\mathrm{halo}=10^{13}$ at $z=0$) may have have a steep decline from $z=5$ to $z=0$ \citep{2020A&A...643A...5D}. {Another caveat is that we are using multiple gas tracers for the kinematics but focus on molecular gas fractions, as ionised gas does not significantly contribute to the disc mass. We discuss this in more detail in Section~\ref{sub.phase}.}

\begin{figure}
\includegraphics[scale=0.5,trim=0.5cm 2.5cm 0cm 0.8cm, clip]{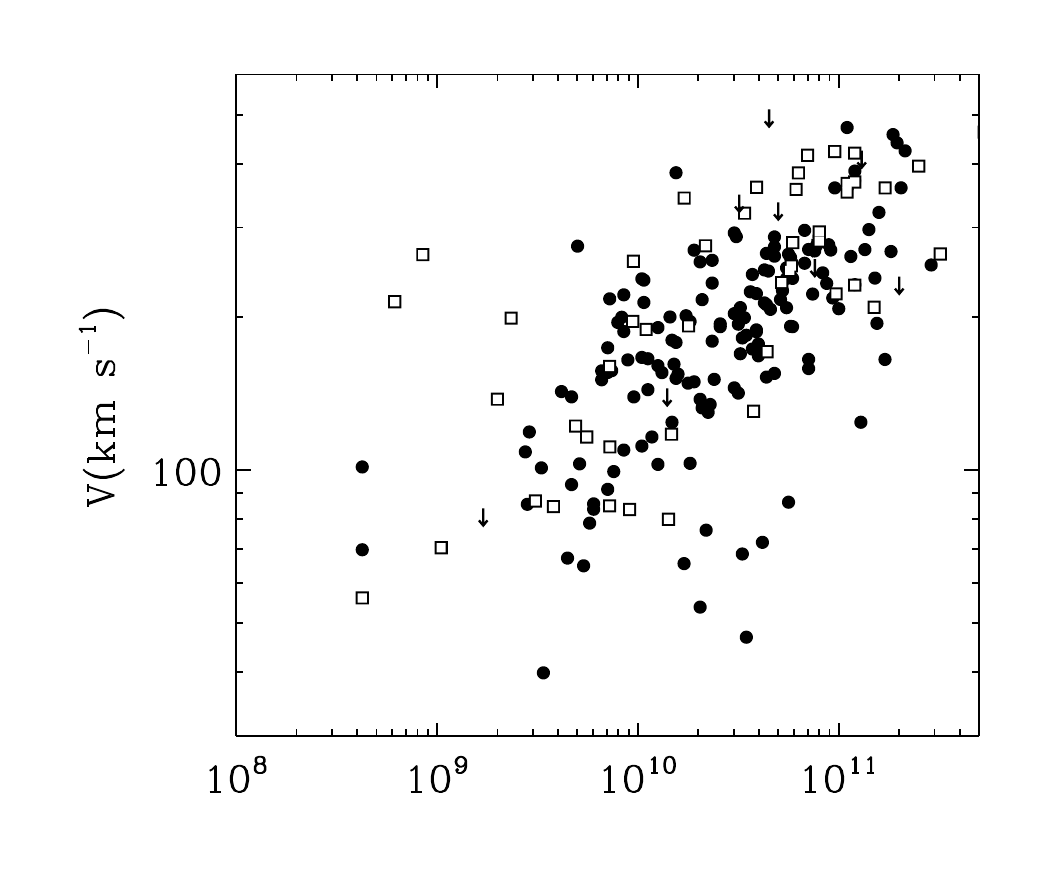}
\includegraphics[scale=0.5,trim=0.5cm 0.6cm 0cm 1cm, clip]{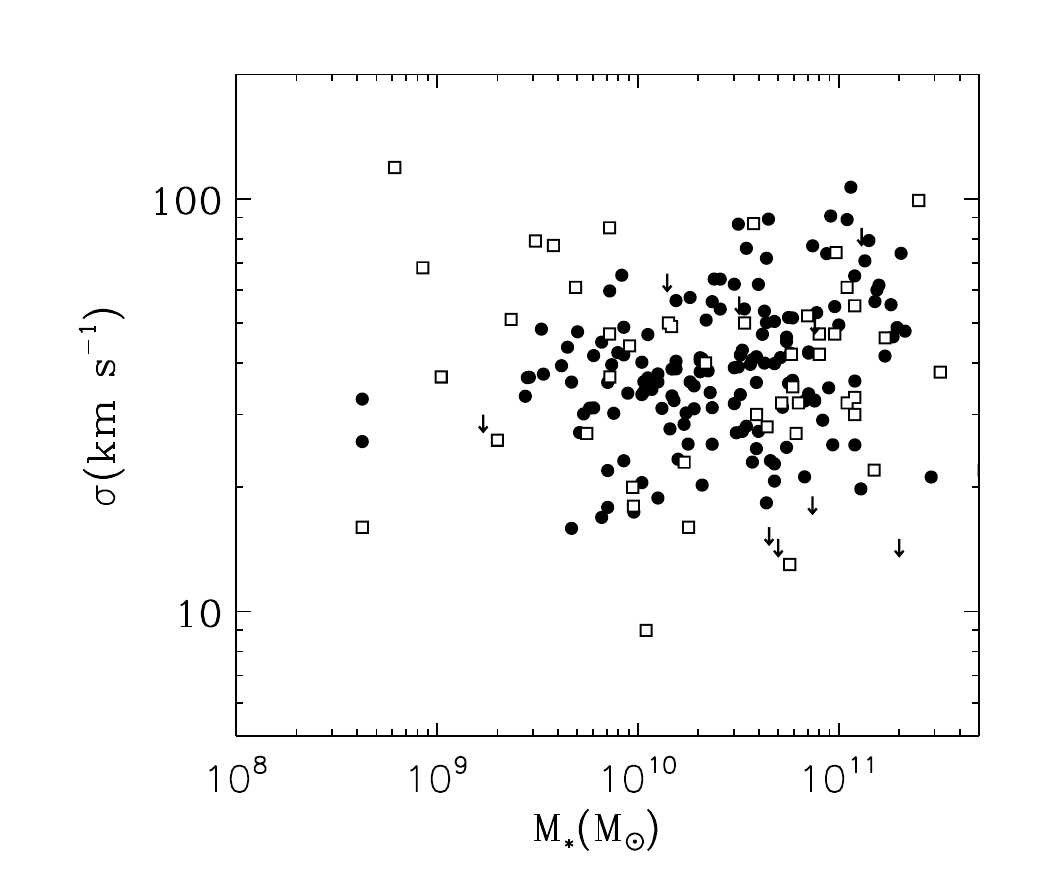}
\caption{Inclination corrected rotational velocity, $V$, and disc velocity dispersion, $\sigma$, as a function of stellar mass, \Mstar, for the data compilation. Ionised gas tracers are shown as black points and the FIR/sub-mm sample is shown with open squares. While there is a clear correlation for $V$-\Mstar (Pearson correlation, $r=0.72$), reflective of the Tully-Fisher relation, there is no correlation for $\sigma$-\Mstar (Pearson correlation, $r=0.07$), however we note here that $\sigma$ from the observations is the line-of-sight velocity dispersion. }
\label{afig.massdep}
\end{figure}

\begin{figure*}
\includegraphics[ scale=0.725, angle=90, trim=8.6cm 0cm 0cm 0.0cm, clip]{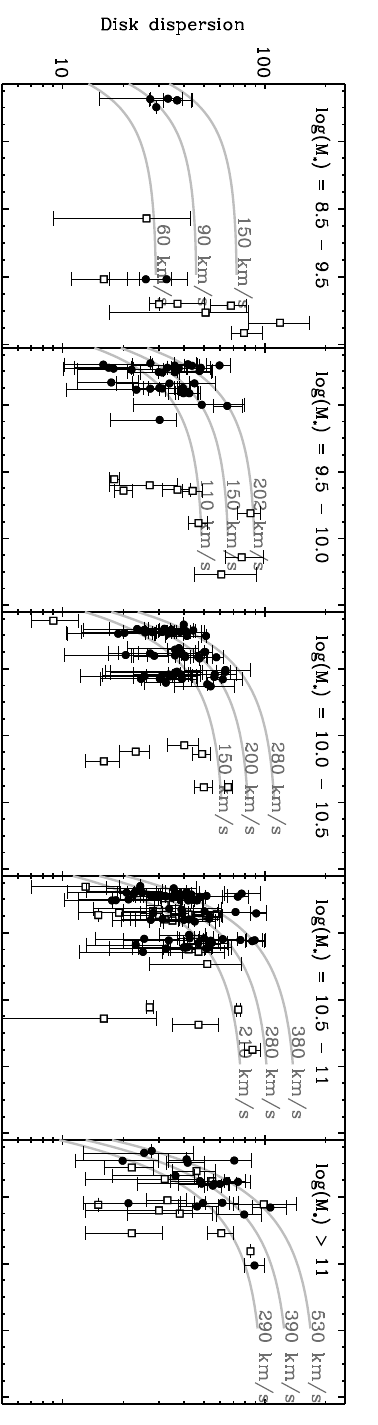}
\includegraphics[ scale=0.725, angle=90, trim=7.6cm 0cm 0.1cm 0.0cm, clip]{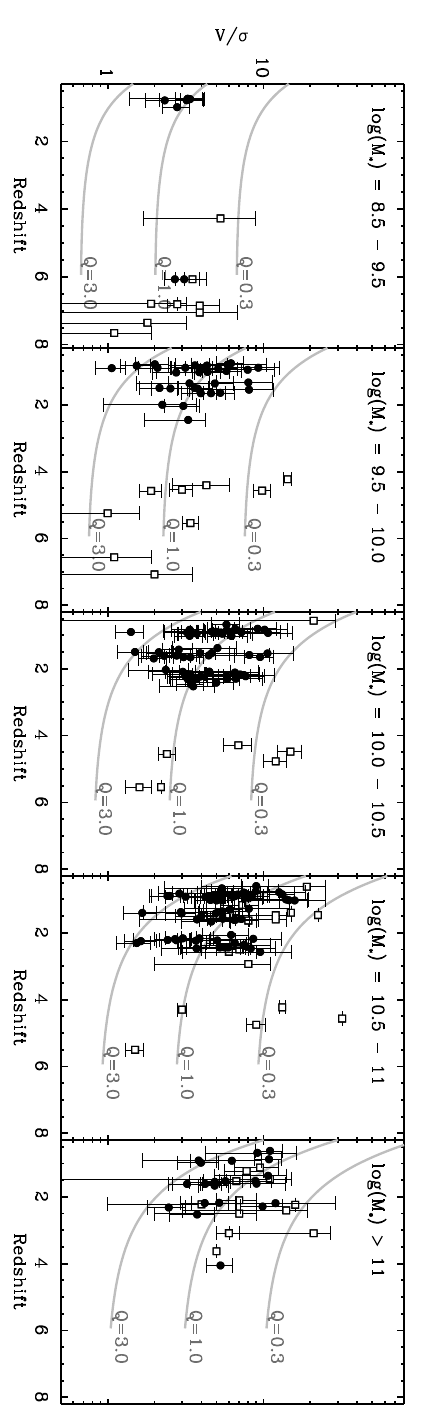}
\caption{Gas velocity dispersion (top) and the ratio or gas rotational velocity to velocity dispersion in stellar mass bins. The data are the same as Fig.~\ref{fig.Tkeyplot}, optical (black circles) and FIR/mm (open squares).  The analytic prescription, described in Section~\ref{sec.models}, is shown for different rotational velocities (top panels) and Toomre $Q$ parameters (bottom). The rotational velocities chosen represent the range expected for each mass bin given the $z\sim1-2$ Tully-Fisher relation and intrinsic scatter defined in equation 2 and Table 2 of \protect\cite{2017ApJ...842..121U} respectively.  
}
\label{fig.Tmass}
\end{figure*}
\subsection{The role of specific star formation rate evolution}
\label{sub.ssfr}
In comparison to depletion time or molecular gas content, SFRs are far easier to estimate for large galaxy populations. However, due to the number of techniques used to estimate SFR and the variety of data quality, it can be difficult to measure the evolution of sSFR consistently across all of cosmic time. Large compilations spanning wide redshift and mass range include \cite{2014ApJS..214...15S}, \cite{2020ApJ...899...58L}, \cite{2021MNRAS.505..540T}, and \cite{2023MNRAS.519.1526P}. The exact shape is influenced by the SFR indicator used, how star-forming and passive galaxies are separated, observational biases etc. (e.g., \citealt{2022ApJ...936..165L}). In the left panels of Fig.~\ref{afig.comp} we show the comparison of different sSFR($z$) parameterisations in four different stellar mass bins. Despite the challenges mentioned, the different parameterisations for sSFR are in good agreement across $10^{9.5}-10^{11}$ \Msun with some minor normalisation differences which are more pronounced in the highest mass bin. The slope of the evolution of \cite{2014ApJ...795..104W}, used in \citetalias{2015ApJ...799..209W}, is marginally steeper than the other parameterisations between $z=0-3$ at all masses. 

For this work we adopt the sSFR($z$) from \cite{2014ApJS..214...15S} for consistency because it was used in deriving the evolution of depletion time and molecular gas mas in \cite{2020ARA&A..58..157T}. The last column of Fig.~\ref{afig.comp} shows how $f_\mathrm{gas}$($z$) and therefore $\sigma$($z$) would change if we use different observationally-derived prescriptions for the evolution of sSFR($z$) and \tdep$(t$) together. There is a significant normalisation difference particularly at low mass (\Mstar$<10^{10}$\Msun) and high redshift ($z>3$). This difference may be due to low number statistics in this regime in deriving both sSFR($z$) and $t_\mathrm{dep}$($z$). The steepness in the $\sigma$($t$) model from \citetalias{2015ApJ...799..209W}, red band in Fig.~\ref{fig.Tkeyplot}, can be seen here as directly related to the use of the sSFR($z$) from \cite{2014ApJ...795..104W}. Interestingly, at $z>3$ the derivations using \cite{2018ApJ...853..179T,2020ARA&A..58..157T} with \cite{2020ApJ...899...58L,2023MNRAS.519.1526P} show a turn-over to lower \fgas($z$,\Mstar), and therefore lower $\sigma$($z$,\Mstar), to higher redshifts.

\subsection{Stellar mass assumptions}
\label{sub.mass}
As discussed in Section~\ref{sub.molgas} and Section~\ref{sub.ssfr} and shown in Fig.~\ref{afig.comp}, the shape and normalisation of the derived model of $\sigma$($z$) is dependent on stellar mass.
For Fig.~\ref{fig.Tkeyplot} we assume an average mass, $\log M_*$[\Msun] $=10.5$, which matches the mean of the full sample. However, given the wide range of \Mstar~covered by the data compilation, we explore the dependence of the data compilation and updated model on \Mstar~in Fig.~\ref{afig.massdep} and Fig.~\ref{fig.Tmass}. The ionised gas sample spans the whole mass range with the highest concentration of galaxies at $\log M_*$[\Msun] $=10-11$. In contrast the mm/FIR sample more uniformly spans the full mass range. It is possible that disc dispersions correlate with stellar mass either directly or indirectly through secondary correlations (e.g., SFR-\Mstar, $V/\sigma$-\Mstar). In Fig.~\ref{afig.massdep} we show the $\sigma$ and $V$ as a function of stellar mass. While there is a clear correlation between $V$ and $M_*$, as expected by the Tully-Fisher relation, there is no strong correlation between $\sigma$ and $M_*$. This has been previously explored in the ionised gas data in \cite{2017ApJ...842..121U} and \cite{2019ApJ...880...48U}.

The expected evolution using Toomre stability arguments is shown with gray lines for appropriate choices of rotational velocity for each mass bin in Fig.~\ref{fig.Tmass}. A steeper population evolution is predicted at high mass than at low mass, specifically with respect to $z=1-4$, as expected from the right panels of Fig.~\ref{afig.comp}. The majority, approximately two thirds, of the ionised gas data overlaps with the model expectations for the full mass range, with better agreement in the lowest ($\log$(\Mstar[\Msun]$)=8.5-9.5$) and highest ($\log$(\Mstar[\Msun]$)>11$) mass bins, although those have the lowest number statistics. In contrast the millimeter data compilation is mostly below the model expectations, except at $\log$(\Mstar[\Msun]$)<10.0$, which may indicate a difference in the kinematics between phases (e.g., \citealt{2019ApJ...880...48U,2021ApJ...909...12G}). We explore the differences in $\sigma$ between gas phases in Section~\ref{sub.phase}.

\subsection{Other model assumptions}
\label{sub.other}
The Toomre model also includes the variables $a$, $V$, and $Q$. For comparison to the literature dataset, a value or range has been assumed. We explore those choices here. 
The exact geometry of the galaxies considered has implications for the assumed value of the constant $a$. In \citetalias{2015ApJ...799..209W} and in this work we assumed $a=\sqrt{2}$. The choice of $a$ can have an effect of up to a factor of 2 on the expectation from equation~\ref{eq.toomre}.

The width of the model band is determined by a range of circular velocities. Due to the relatively tight correlation between mass and velocity (e.g., \citealt{1977A&A....54..661T,1985ApJS...58...67T,2000ApJ...533L..99M}) the choice of $V$ should reflect the appropriate range based on the range of stellar masses for the dataset. For Fig.~\ref{fig.Tkeyplot} we use $100-250$ km s$^{-1}$.  The top panels of Fig.~\ref{fig.Tmass} show three curves reflecting appropriate rotational velocities assuming minimal evolution of the stellar Tully-Fisher relation from $z\sim1-8$ using \cite{2017ApJ...842..121U} in mass bins of 0.5 dex. The majority of data fall between the expected rotational velocities but a significant fraction of data, particularly the mm/FIR measurement, across all redshifts scatter to lower dispersions. The majority of mm/FIR measurements (white squares) that have measured dispersions lower than the expectation, shown by the lowest gray line, actually have corresponding rotational velocities that are much higher than expected by the model (up to 500 km s$^{-1}$), indicative of a high $V/\sigma$ and/or much lower $Q$. 

The measured $V/\sigma$ are explored in the bottom panels of Fig.~\ref{fig.Tmass} with model lines for $Q=0.3,1.0,3.0$. In comparison to Fig.~\ref{fig.Tkeyplot} and the top panels of Fig.~\ref{fig.Tmass} these panels take into account an extra observable, $V$. A wide range of $V/\sigma$ values, from $\sim1-20$, are seen in the literature across all stellar masses, consistent with model expectations of $Q\sim1$ on average. The increasing model values of $V/\sigma$ with \Mstar~reflects the decreasing fraction of \fgas~with \Mstar. In the data, $V$ is strongly correlated with \Mstar~(as expected by the Tully-Fisher relation; Fig.~\ref{afig.massdep}) while $\sigma$ shows no correlation with \Mstar. 

For thin disks, $Q=1$ is the commonly used critical value to define marginal stability \citep{2008gady.book.....B}. This assumes a single phase infinitesimally thin disc. However, it is well know that galactic discs are composed of multi-phase gaseous components probed by HI and H$_2$, HII regions and young stars, and old stars. Theoretical works have shown that the global stability of the multi-phase disc can differ from the stability of any one individual phase. In particular, some simulations \citep{2021MNRAS.508..352R} show that locally HI discs can have higher Toomre stability parameters of $Q_\mathrm{HI}\gtrsim10$, molecular discs have $Q_\mathrm{H_2}\sim10$ and stellar discs have $Q_\mathrm{*}\sim1-3$ \citep{2014ApJ...785...43W}. In contrast, Fig.~\ref{fig.Tmass} (bottom) shows that the cooler gas tracers (mm/FIR lines) are consistent with $Q\lesssim1$. A low $Q$ parameter, $Q\sim0.3$, has been noted in high redshift clumpy cool gas disks \citep{2024arXiv240218543F}. Assuming different combined values of $Q$ would change the normalisation of the expected evolution of disc velocity dispersions. 

An additional complication is that the Toomre arguments used here assume a infinitesimally thin disc, which is an unjustified assumption for the majority of the data in this compilation \citep{2006ApJ...650..644E,2014ApJ...792L...6V,2017ApJ...847...14E,2024ApJ...960L..10L, 2024arXiv240915909T}\footnote{An additional complication of the Toomre formalism used here is whether the scale that the dispersion is measured at is above or below the scale height of the disc \citep{2021MNRAS.508..352R}. Because the observations utilised can not independently measure these two parameters we are unable to consider a more precise formalism}. Theoretical derivations of disc stability criteria of thick discs imply a lower value of $Q_\mathrm{crit}\sim0.7$ {\citep{2002ApJ...581.1080K, 2024A&A...687A.115B}.} The formalism introduced by \cite{2013MNRAS.433.1389R} includes both a term that accounts for the stabilisation effect due to finite thickness as well as a weights for each component. The reduction in $Q_\mathrm{crit}$ would result in a lower expectation for $\sigma$ bringing the gray band in Fig.~\ref{fig.Tkeyplot} closer to the lower envelope of measurements.

\section{Discussion}
\label{sec.discussion}

\subsection{Shape of the evolution} 
While ionised gas results indicated a steady evolution between $z=0.5-2.5$ (e.g., \citealt{2017ApJ...843...46S, 2019ApJ...886..124W}, recent work has claimed no evolution in $V$/$\sigma$ between $z=0.5-3.5$ \citep{2023AA...679A.129R}, meanwhile other works have suggested discs at $z\sim4-8$ are dynamically colder than expected \citep{2020Natur.584..201R,2023AA...672A.106L}. The expected shape of the evolution of dispersion at fixed mass is most strongly tied to the co-evolution of gas fractions and sSFR at fixed mass (Fig.~\ref{afig.comp}, Fig.~\ref{fig.Tmass}). As a result, the expectation of the shape of evolution beyond $z\sim4$ is still uncertain with fewer measurement constraints on gas fraction across a wide mass range. Using updated prescriptions for the evolution of depletion time and specific star formation rates, the expected evolution of disc dispersions would flatten beyond $z\sim1$, with a mild dependence on stellar mass. Fig.~\ref{fig.Tmass} shows that a smaller difference in measured disc dispersions is expected as a function of cosmic time for low mass systems compared to high mass systems. This could be linked to the `disc settling' scenario \citep{2012ApJ...758..106K,2024arXiv240915909T} in which galaxies generally settle as they become more massive and from $z\simeq1$ experience less mergers and accretion with cosmic time (e.g., \citealt{, 2019ApJ...886..124W}). { A similar shaped evolution has been derived using a purely feedback driven model \citep{2024A&A...689A.273R} with dependencies on disc scale height, molecular gas mass and total star-formation rates.}

Fig.~\ref{fig.Tkeyplot} and Fig.~\ref{fig.Tmass} reveal that most literature results of $\sigma$ at kpc-scales at $z>4$ are consistent with theoretical predictions from marginal stability arguments, with considerable scatter towards lower dispersions. This is best seen in Fig.~\ref{fig.Tmass}, where it becomes clear that the gas phase used for measurement may also have an effect on the measured dispersions and stability, as discussed in Section~\ref{sub.mass} and explored more in Section~\ref{sub.phase}.

The evolution at fixed mass has a self-similar shape to the recently measured evolution of disc thickness for low-mass edge-on galaxies in \textit{JWST} images ($\log$(\Mstar[\Msun]$)=8.5-10.5$) from $z\sim5-0$ \citep{2024ApJ...960L..10L}. In that study, a flat evolution is measured between $z\sim3-1$ of $\sim0.4$ kpc with $1\sigma$ scatter of 0.15 kpc and a decline to $\sim0.2$ kpc at $z=0$ (but see \citealt{2024arXiv240915909T}). Together, these results suggest, that at least for low masses, discs can form in a thick configuration at early times or become thick quickly. Dynamical `heating' likely occurs from a number of processes after formation contributing to the scatter in the thickness of stellar discs. Once thick,  subsequent minor mergers or secular processes become dynamically inefficient with only major mergers possible to remove the existing thick disk.

The simple model does not show the evolution of individual galaxies but likely captures the population average at various redshifts due to minimal evolution of the sSFR and \fgas~relations beyond $z\simeq1$. Pathways of individual galaxies should be varying significantly in diverse ways due to their varied star formation and merger histories.

\subsection{Scatter}
\label{sec.scatter}
While the global evolution of mass averaged measurements can be {modelled} using Toomre stability theory, the scatter at any given epoch reflects the combination of internal physics driving the pc-scale motions in the ISM as well as systematic uncertainties (resolution effects, tracer, measurement methodology). What physical mechanisms {drive and maintain disc-scale turbulence} across cosmic time remains an elusive problem due to the combination of systematic uncertainties (e.g., \citealt{2018MNRAS.474.5076J,2019ApJ...880...48U}). In this work we find no direct dependency on stellar mass, though secondary correlations could exist due to the connection between stellar mass and gas fractions or more massive galaxies becoming more stable (e.g. $Q>1$).

How the scatter of kinematic measurements relates to the molecular gas reservoir through measurements of \Mgas, \fgas, and/or \tdep~has been particularly observationally challenging. Large statistical studies of multi-phase tracers have not been obtained. These limitations motivate using the large but heterogeneous literature compilation presented in Section~\ref{sec.data}. In the following sections, we utilise the smaller sub-sample of \Ngalgas~galaxies with both kinematic and gas reservoir estimates.

\begin{figure*}
\includegraphics[ scale=0.375,trim=1.0cm 0.0cm 0cm 0.0cm, clip]{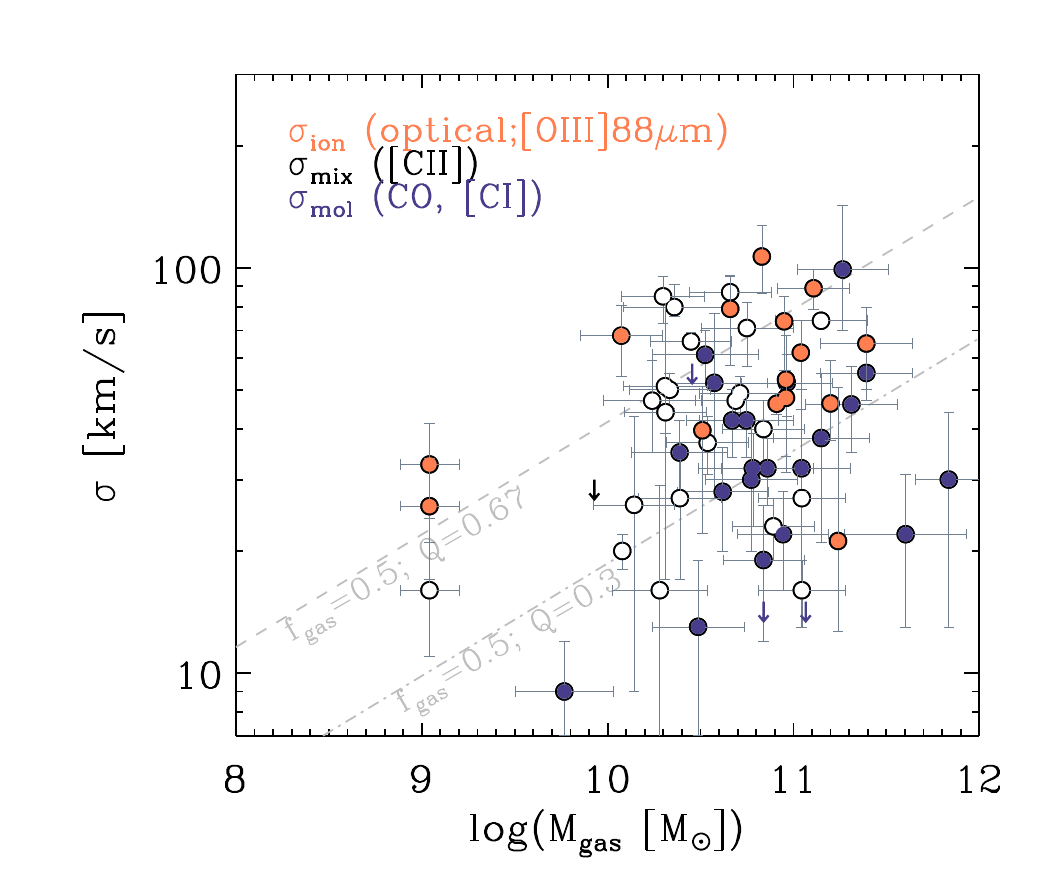}
\includegraphics[ scale=0.375,trim=2.2cm 0.0cm 0cm 0.0cm, clip]{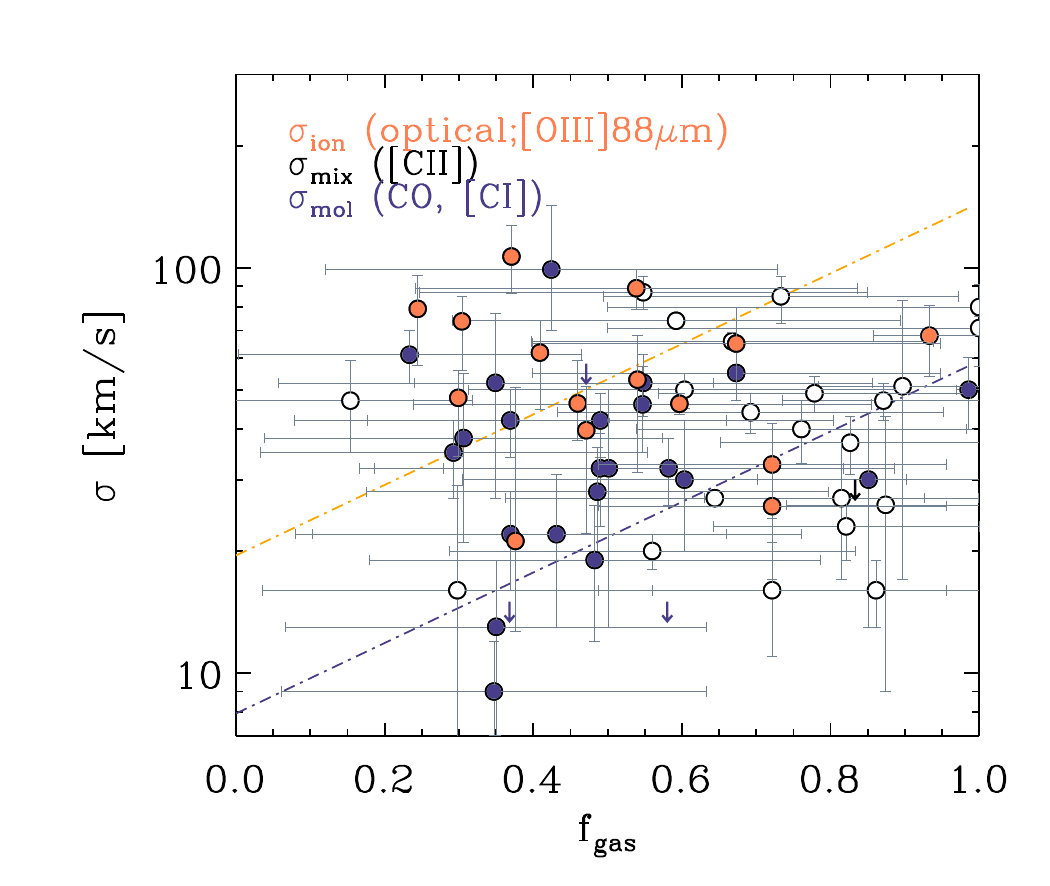}
\includegraphics[ scale=0.375,trim=2.2cm 0.0cm 1cm 0.0cm, clip]{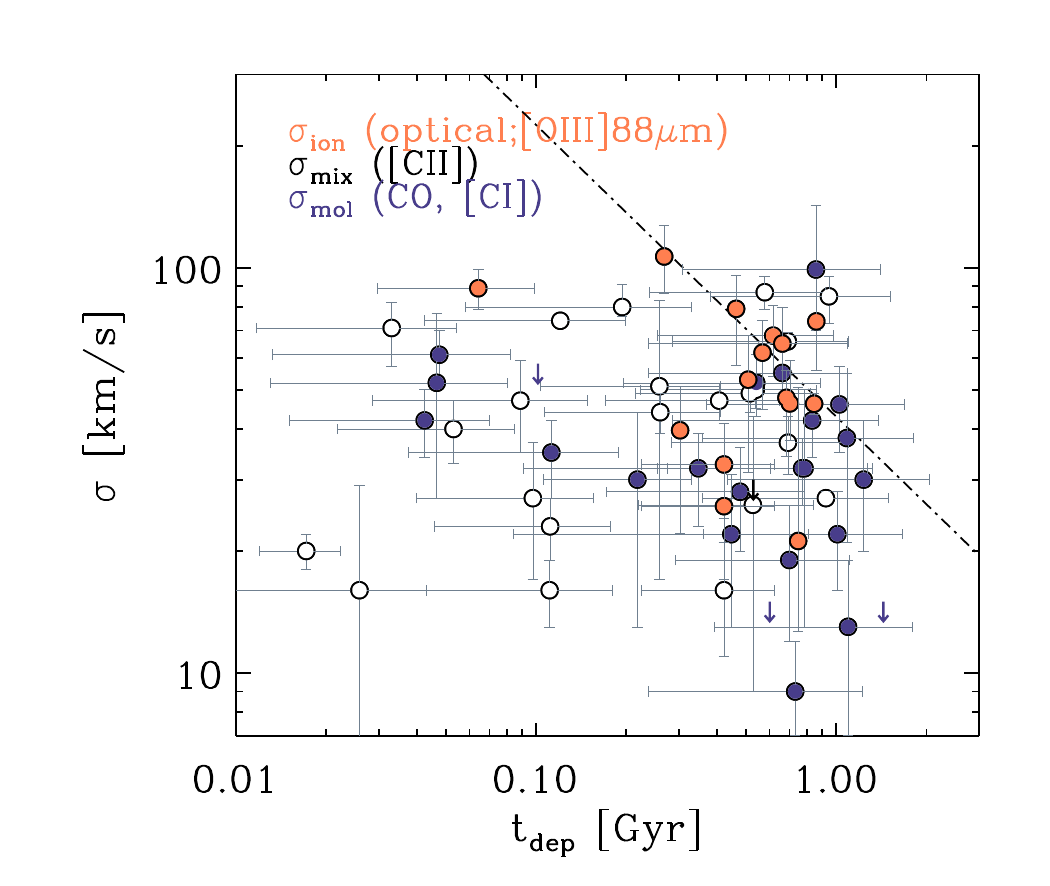}
\caption{Velocity dispersion as a function of molecular gas properties: \Mgas, \fgas, \tdep~from left to right for galaxies with a molecular gas mass, stellar mass, and SFR measurement. Galaxies are color coded by gas phase. Orange points represent galaxies with kinematic measurements of rest-frame optical emission lines (H$\alpha$, \oiii) and the \oiii $88\mu$m line. Dark blue points represent galaxies with kinematic measurements of cooler gas tracers including CO transitions and \ci370$\mu$m, \ci609$\mu$m. Open white symbols represent galaxies with kinematic measurements from \cii$158\mu$m a mixed gas phase tracer. Dashed lines in the left panel show equation~\ref{eq.toomre2} assuming the \citep{2017ApJ...842..121U} Tully-Fisher relation, a molecular gas fraction of 50\% and $Q=0.3,0.67$. The dot-dashed lines in the middle panel show fitted relations from \protect\cite{2021ApJ...909...12G}. The dot-dash line in the right panel shows the fit to local data from \protect\cite{2019ApJ...870...46F} consistent with the multi-freefall turbulence models of \protect\cite{2015ApJ...806L..36S}. Arrows represent upper limits of the dispersion. Errors show propagated uncertainties assuming a 0.3 dex error on stellar mass measurements and a 30\% error on SFR measurements. }
\label{fig.gas1}
\end{figure*}

\begin{figure*}
\includegraphics[ scale=0.365,trim=1.0cm 0.0cm 0cm 0.0cm, clip]{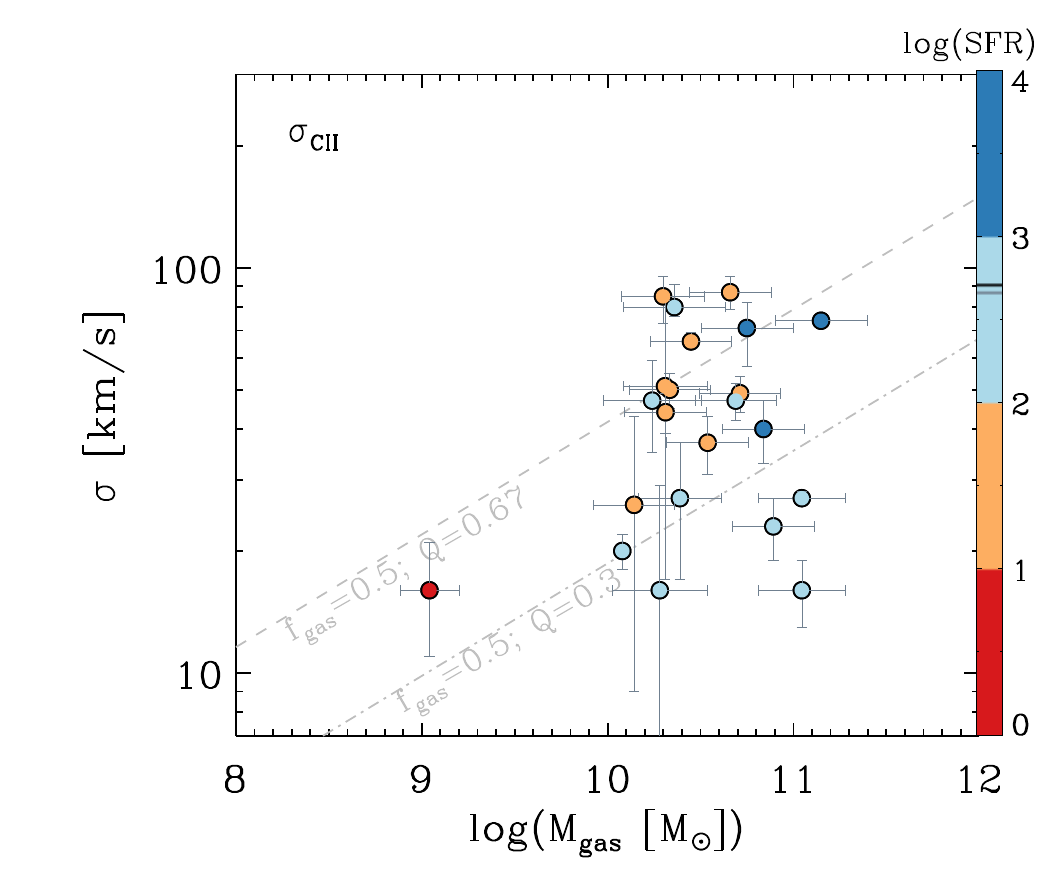}
\includegraphics[ scale=0.365,trim=2.2cm 0.0cm 0cm 0.0cm, clip]{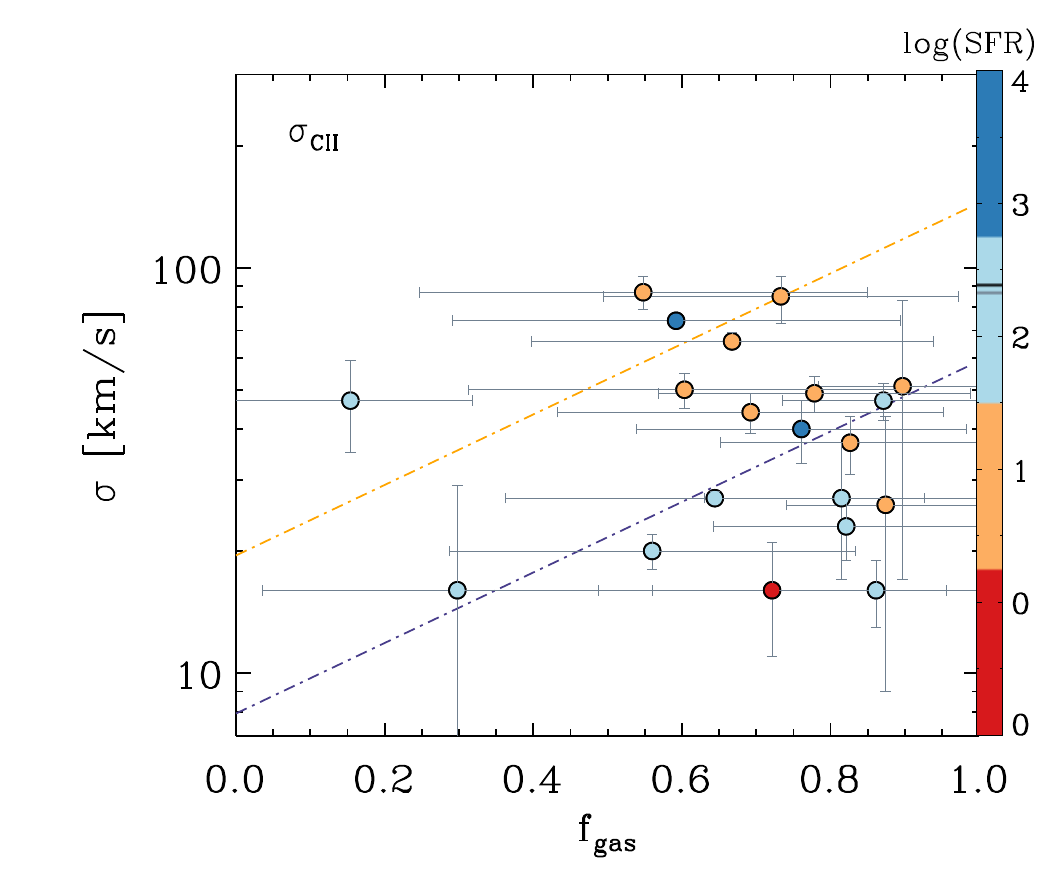}
\includegraphics[ scale=0.365,trim=2.2cm 0.0cm 0cm 0.0cm, clip]{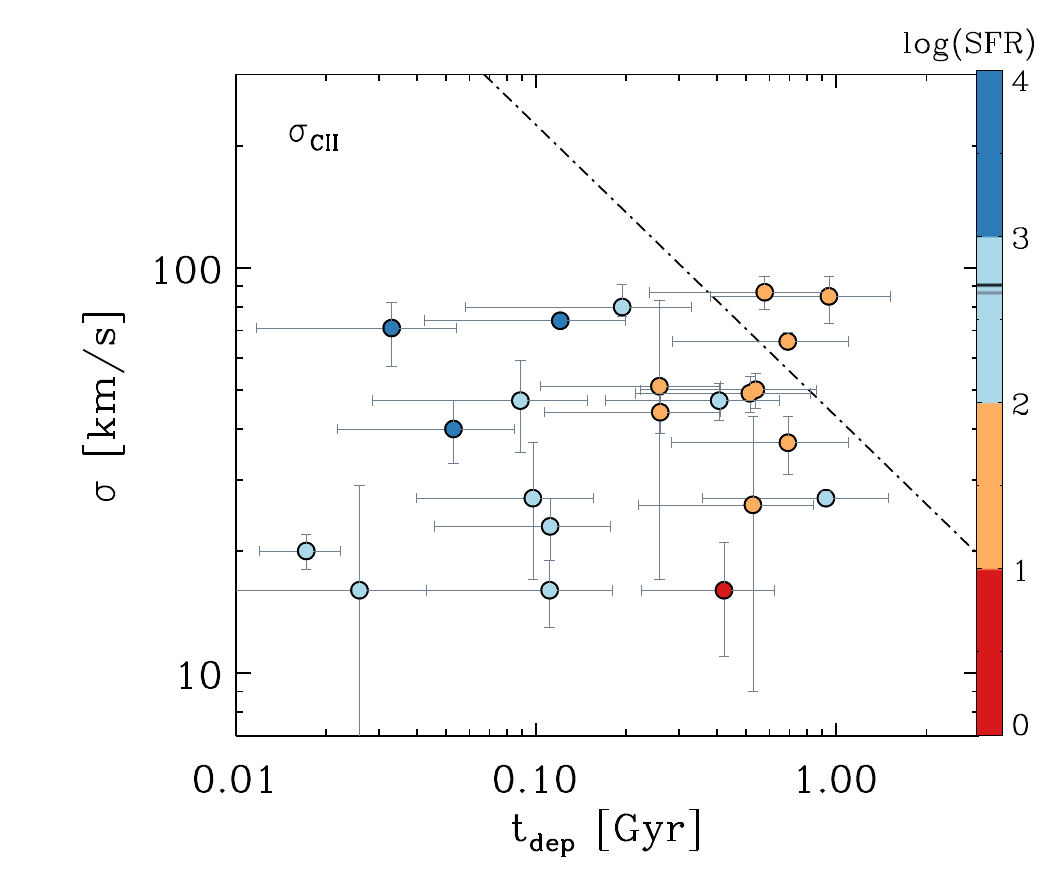}
\caption{Same as Fig.~\ref{fig.gas1} but showing only kinematic measurements derived from \cii. Color coding represents measured star formation rates as denoted by the color bars. Star formation rates are taken from multiple tracers depending on the observations available as described in Section~\ref{sec.data}. }
\label{fig.gas2}
\end{figure*}

\subsubsection{Offset between gas phases}
\label{sub.phase}
Using the homogenised \Mgas~values described in Section~\ref{sec.data}, we directly compare the velocity dispersion with measurements of \Mgas, \fgas=\Mgas/(\Mgas+\Mstar), and \tdep=\Mgas/SFR independent of redshift.
Fig.~\ref{fig.gas1} (left) shows the dispersion as a function of \Mgas~color-coded to indicate which gas phase is being used to trace the kinematics. For ionised gas tracers we include data from rest-optical lines (\halpha, \oiii5007\AA) and rest FIR line \oiii88$\mu m$ (which traces HII regions around young stars; \citealt{1975ApJ...202L..31W,1985ApJS...57..349R}). For molecular gas tracers we include data from CO transitions as well as \ci370$\mu$m, \ci609$\mu$m. 
There has been both theoretical and observational evidence that \ci traces molecular clouds similar to low J CO transitions (e.g., \citealt{2002ApJS..139..467I, 2013MNRAS.435.1493A, 2019MNRAS.486.4622C}). We do not classify \cii158$\mu$m as either ionised or molecular gas tracer as it is found in regions of ionized, molecular, and neutral gas over a large range of temperatures (e.g., \citealt{2012ApJS..203...13G,2014A&A...570A.121P}). 

From the left panel of Fig.~\ref{fig.gas1} we see a separation between ionised and molecular gas dispersions at fixed \Mgas, such that ionised gas tracers (orange) cluster to higher dispersions than molecular gas tracers (blue). Expectations from Toomre stability theory are included for $Q=0.3$ and $Q=0.67$, assuming a 50\% molecular gas fraction and that the Tully-Fisher relation holds and does not evolve significantly at high redshifts. While the data sample is limited to high gas masses, the data are not inconsistent with the model expectation.  As expected from Fig~\ref{fig.Tmass}, the molecular gas dispersions align better with the $Q=0.3$ model while the ionised gas dispersions align better with the $Q=0.67$ model. These results indicate that high dispersions of $\sim$50 \kms~can be reached in the molecular phase but only in highly unstable disks with large gas masses of $>10^{11}$\Msun. Ionised gas dispersions are higher by a factor of $\sim2$ on average at fixed gas mass. This offset is comparable to a similar offset seen in $\sigma$, for fixed gas fraction, for a compilation of local analogs of galaxies at cosmic noon (high SFRs, higher velocity dispersions) that have measurements for the same sources in \textit{both} CO and \halpha~\citep{2021ApJ...909...12G}. We do not see as clear of an offset when considering $\sigma$ as a function of \fgas~as shown in the middle panel of Fig.~\ref{fig.gas1}. It is possible this reflects the uncertainty in the measurements (e.g. \Mstar) or could imply that the gas reservoir is more fundamental in setting the dispersion. 

Some of the offset seen may result from the different methods typically used to measure dispersion across samples. As shown in Fig.~\ref{fig.Tkeyplot}, 3D-Barolo is favoured for studies of mm/FIR, while other methods are favoured for optical data. \cite{2024arXiv241107312L} show that in low S/N data dispersions can be underestimated in the outskirts using non-parametric codes. Further studies are required to measure the magnitude of this effect in the current data. 

At $z=0.6-2.7$ the scatter in rest-frame optical dispersions was explored with relation to \Mgas~and \fgas~\citep{2019ApJ...880...48U}, with the molecular gas properties derived from scaling relations \citep{2020ARA&A..58..157T}. The correlations between $\sigma$-\Mgas~and $\sigma$-SFR were shown to have equal Spearman rank coefficient ($\rho_\mathrm{s}=0.38$) and significance ($\sigma_\rho=4.6$) with $\sigma$-\fgas~showing only marginal significance ($\rho_\mathrm{s}=0.12$; $\sigma_\rho=1.4$). However, when the evolution of these parameters were taken into account the correlations became less significant. If we overlay the ionised gas $\sigma$ and derived \Mgas~values from \cite{2019ApJ...880...48U} they overlap with the orange data points in the left panel of Fig.~\ref{fig.gas1} but with higher scatter, likely induced by the derivation of \Mgas~from scaling relations. At higher redshift, $z=4-7$,  \cite{2023AA...673A.153P} compile a sample of 36 galaxies, also included in this work, and find no correlation with \fgas, suggesting a more direct dependence on dust mass or stellar mass.

Offsets in the dispersion of the warm ionised medium and cold neutral medium have also been seen in idealised ISM simulations \citep{2023MNRAS.522.1843R} as a function of \sfrd~due to stellar feedback (a factor of $2.2\times$, consistent with the offset seen by observations in \citealt{2021ApJ...909...12G} and in Fig.~\ref{fig.gas1}). {An offset of $\sim3\times$ was identified in a larger sample of highly star-forming galaxies over $0<z<5$ \citep{2024A&A...689A.273R}.} Together these results {are suggestive of a coexistence of molecular and ionized gas discs with unique stability criteria}. However, it does not necessarily follow that all disks are born with a `cold' molecular disk. Differences in gas phase are also seen in isolated disc and zoom-in simulations \citep{2022MNRAS.514..480E,2024A&A...685A..72K} with $\sigma_\mathrm{molecular}<\sigma_\mathrm{ionised}$. In these simulations, gas-rich discs (\fgas$\sim50$\%) are able to reach levels of molecular gas turbulence of up to 50 \kms~(also seen in Fig.~\ref{fig.gas1}) without any stellar feedback. In these isolated disc simulations and in cosmological simulations \citep{2020MNRAS.496.1620O}, stellar feedback is responsible for the larger ionised gas dispersions or larger scatter, similar to the ISM slab simulations \citep{2023MNRAS.522.1843R}, while molecular gas turbulence is more closely linked to how galactic discs regulate their gravitational stability. Therefore the commonly measured ionised gas disc dispersion is dependent both on how turbulent the molecular gas was that the young stars were born from as well as the amount of star-formation feedback.

A weak trend is also seen when considering the relationship between velocity dispersion and depletion time (Fig.~\ref{fig.gas1}; right). The ionised and molecular gas measurements form overlapping populations close to the expectations from multi-freefall turbulence models from \cite{2015ApJ...806L..36S} and explored in \cite{2019ApJ...870...46F} using local highly star-forming galaxies.  Other models based on feedback-regulated star-formation predict a similar relationship but with a different slope. The data and errors do not justify a distinction between models. While the data compilation presented here does offer some insights into the turbulent nature of different components of the ISM, a dedicated program exploring multiple ISM tracers in the same galaxies is needed to confirm these results.

\subsubsection{Combining \cii measurements together with other tracers}

Due to the brightness of the line, \ciins, has become a popular tracer to explore $z>4$ galaxy kinematics \citep{2020Natur.581..269N,2020Natur.584..201R,2021Sci...372.1201T,2021Sci...371..713L,2021MNRAS.507.3952R}, SFRs (e.g., \citealt{2014A&A...570A.121P,2015ApJ...800....1H, 2024MNRAS.528..499L}), and gas masses (e.g., \citealt{2018MNRAS.481.1976Z, 2020A&A...643A...5D}). The relatively low dispersion values measured with \ciins, indicating dynamically `cold' discs, at $z>4$ \citep{2020Natur.581..269N,2020Natur.584..201R} are surprising given the more turbulent ionised gas results at $z\sim1-3$ \citep{2009ApJ...697.2057L, 2009ApJ...706.1364F,2012ApJ...758..106K,2015ApJ...799..209W}. A possible explanation has been that stars are born out of low-dispersion material, traced by the \ciins, and star-formation driven feedback increases the dispersion, which is preferentially seen by the ionised gas measurements (e.g., \citealt{2024A&A...685A..72K}). In this data compilation, the kinematic measurements from \cii span the range of ionised and molecular gas sequences in Fig.~\ref{fig.gas1} with large scatter. Below we explore possible explanations including physical reasons and measurement uncertainty.

In Fig.~\ref{fig.gas2} we isolate galaxies with kinematic measurements from \ciins, and explore the possible role of a SFR dependence on the phase of gas probed by \cii and thus reflected in the kinematics. The left panel of Fig.~\ref{fig.gas2} shows that, at fixed \Mgas, disc dispersion is lower on average for galaxies with higher SFRs for the \cii kinematic sample. This is contrary to expectations from arguments of star-formation driven turbulence seen in both theory \citep{2013MNRAS.433.1970F,2016MNRAS.458.1671K,2018MNRAS.477.2716K} simulations \citep{2019MNRAS.482.5125H,2020MNRAS.496.1620O,2022MNRAS.514..480E,2023MNRAS.524.4346J} and observations (e.g., \citealt{2009ApJ...699.1660L,Green:2010fk,2019ApJ...880...48U}). The galaxies with SFRs $<100$ \sfrunits~are on the upper envelope of the $\sigma-$\Mgas~trend seen in Fig.~\ref{fig.gas1}. If the result of a kinematic offset in phase discussed above is robust (noting the caveats in Section~\ref{sub.mmobs}), and not dependent on stellar mass (Fig.~\ref{afig.massdep}), then these results indicate that \cii traces a higher fraction of molecular gas in galaxies with high SFRs and primarily traces the warm ionised phase when SFRs are low/average. 

This inference is consistent with some theoretical works and cosmological simulations that suggest molecular gas dominates the [C II] emission at high SFRs (e.g., $>20$ \sfrunits) or SFR densities (e.g., $\Sigma_\mathrm{SFR}>-0.5$ \sfrunits~kpc$^{-2}$), while atomic gas or gas in photo-dissociated regions (PDRs) takes over at lower SFRs and SFR densities \citep{2015ApJ...814...76O}. The change with SFR density is suggestive that \cii traces mostly molecular gas in high-density/pressure regions, and otherwise traces the atomic/PDR gas phase \citep{2017MNRAS.467...50N}. We do not explore the relation with SFR density in the data compilation due to large uncertainties in size measurements in the high redshift data. 

Simulations (e.g., \citealt{2024MNRAS.528..499L, 2024ApJ...965..179G}) and observations (e.g., \citealt{2014A&A...568A..62D, 2015ApJ...800....1H, 2017ApJ...845...96C}) have explored the possible metallicity dependence of the \ciins$-$SFR relation which could also play a role in how best to interpret which phase is dominating \cii kinematics.  \cite{2024ApJ...965..179G} find that the fraction of \cii emission originating in ionized gas increases with galaxy metallicity, consistent with some observations \citep{2017ApJ...845...96C, 2020A&A...643A.141M}. In contrast, in the FIRE simulations \citep{2024MNRAS.528..499L} the fractional contributions of \cii emission from different phases only depends strongly on metallicity above solar metallicities but shows a stronger dependence on depletion time at higher redshifts where metallicities are low. The very low depletion times of the high SFR sample in Fig.~\ref{fig.gas2} (right) are associated to 20-40\% molecular fractions of the \cii in the FIRE simulations. 

We are unable to explore the relationship directly with metallicity as most sources with \cii detections do not have metallicity measurements as well. However, focusing on galaxies at a fixed gas mass of \Mgas$=10^{10}-10^{11}$\Msun~in the left panel of Fig.~\ref{fig.gas2} and assuming either the fundamental mass metallicity relation \citep{2010MNRAS.408.2115M} or the gas mass metallicity relation \citep{2016MNRAS.455.1156B} the galaxies with high star-formation rates (low dispersions) should have lower metallicities. This goes in the direction expected from \cii studies that suggest molecular gas is traced well by \cii in low metallicity (low dust) regions \citep{2020A&A...643A.141M}. 
 
An alternative explanation for the spread in \cii measurements is that the gas masses derived from dust continuum and rest-frame FIR lines for the highly star-forming galaxies are overestimated by roughly an order of magnitude. \cite{2020A&A...634L..14C} study the dust temperature of GN20, a galaxy within our sample \citep{2024MNRAS.533.4287U} finding that, assuming a constant gas to dust ratio, a dust temperature of 25 K verses 50 K could result in a 7$\times$ over-estimate of the gas mass. A reduction in the gas masses of the high SFR galaxies would bring them more in line with the lower SFR galaxies in Fig.~\ref{fig.gas2} (left and middle). However, a reduction in \Mgas~for the high SFR galaxies would also lead to further reduction in the depletion time, in contrast to expectations from theory and observations (right).  

The \cii measurements at $z>4$ are a key tracer available to measure gas dynamics with high accuracy due to the brightness of the line. While \textit{JWST} will increasingly be able to explore the ionised gas dispersions (e.g., \citealt{2024AA...684A..87D,2023arXiv231006887N,2024MNRAS.533.4287U, 2024arXiv240808350B, 2025arXiv250321863D}) the NIRSPEC IFU and microshutters have a limited spectral resolution, uncertain line spread function \citep{2024AA...684A..87D}, and difficult PSF \citep{2024NatAs...8.1443D}. Therefore it is critical for dynamical studies to better understand the origin of \cii in individual sources for comparison to existing literature.

\section{Conclusions}
We present a literature compilation of molecular and ionised gas kinematics at $z=0.5-8$ of \Ngal galaxies hosting rotation. The sample spans four orders of magnitude in stellar mass, four orders of magnitude in star formation rates, and three orders of magnitude in molecular gas mass. The data come from ground and spaced based optical and near-infrared integral field spectrograph observations, ground-based millimeter interferometer observations, and new $JWST$ NIRSPEC observations. We find that kinematic measurements from far-infrared lines (traced by velocity dispersion, $\sigma$, and rotational support, $V$/$\sigma$) at $z>4$ show significant scatter at fixed redshift comparable to ionised gas results at $z<4$ likely dominated by the heterogeneous nature of the sample in data quality and galaxy properties. 

Using the large literature compilation we explore the evolution in $\sigma$, $V$/$\sigma$, and Toomre stability, $Q$, from $z=0.5-8$ finding no evolution in $\sigma$ within the errors between $z\sim1$ and $z\sim8$. This is consistent with simplified single component Toomre stability arguments in which the average evolution of gas dispersion at fixed mass is not expected to evolve significantly. While previous work, \cite{2015ApJ...799..209W}, presented a model that indicates a continually increasing dispersion with redshift if extrapolated, we present an updated model that predicts little evolution between $z=6$ and $z=2$ except in the highest mass bin ($\log M_*$[\Msun$]>11$). This is confirmed, with considerable scatter, by the data compilation out to $z\sim8$. We explore the effects on the expected average dispersion evolutions from different empirically derived evolution for gas fractions, depletion time, and sSFRs.

We identify a $\sim2\times$ offset between velocity dispersion measured from molecular gas (as measured from CO, \ci, and \oi) and ionised gas (\halpha, \oiii) at a fixed molecular gas mass, \Mgas, indicating a combination of physical processes driving a cooler molecular disc surrounded by a more turbulent ionised disc, consistent with previous literature compilations at lower redshifts and some zoom simulations. However, high values of dispersion in molecular gas discs ($\sim50$ \kms) are measured at high gas masses following expectations of a correlation between $\sigma$ and \Mgas. 

Kinematic measurements using \cii do not follow either the ionised gas or molecular gas expectations showing lower $\sigma$ for higher SFRs at a fixed \Mgas. This is likely due to the the fact that \cii emission can originate from different phases of the ISM with galaxies having different relative contributions from e.g., photo-dissociated regions, neutral gas, CO-dark molecular gas, and ionised gas. When split in SFR bins, the \cii sample behaves as expected with the high SFR (low metallicity) having low dispersions at a fixed \Mgas.

To further pick out the physical meaning underlying the scatter of velocity dispersion in disc galaxies, large kinematic surveys with reliable molecular gas tracers, SFR indicators, and kinematics are needed. However, the underlying driver or maintenance mode of turbulence likely acts on much smaller scales requiring a `PHANGS-like' survey at higher redshift with highly resolved ionised and multiple molecular gas tracers in the \textit{same} galaxies allowing an investigation of $\Sigma_\mathrm{SFR}$, residual velocities, and winds, well below the kpc-scale. For ionised gas, this will only be feasible with 30m telescopes at $z>1$ and the new MAVIS IFU at $z<1$, and for molecular gas with upgraded sub-millimetre facilities.

\section*{Acknowledgements}

EW is grateful for the encouragement and feedback from P. Sharda, K. Glazebrook, S. Brough, M. Kaasinen. 
EW acknowledges support by the Australian Research Council Centre of Excellence for All Sky Astrophysics in 3 Dimensions (ASTRO 3D), through project number CE170100013. EW acknowledges the support of the Kavli Institute for Cosmology, Cambridge Visitor scheme. 
H\"U acknowledges funding by the European Union (ERC APEX, 101164796). Views and opinions expressed are however those of the authors only and do not necessarily reflect those of the European Union or the European Research Council Executive Agency. Neither the European Union nor the granting authority can be held responsible for them.

\section*{Data Availability}

This paper takes advantage of data literature sources as described in Section~\ref{sec.data}. Parameters have been adjusted from originally published sources to homogenise techniques where feasible. The data compilation for this analysis will be provided online.



\bibliographystyle{mnras}
\bibliography{bibdesk}




\appendix
\section{Data compilation}
\label{app.datacomp}

\begin{figure}
\includegraphics[ scale=0.5,trim=1.0cm 0cm 0cm 0cm, clip]{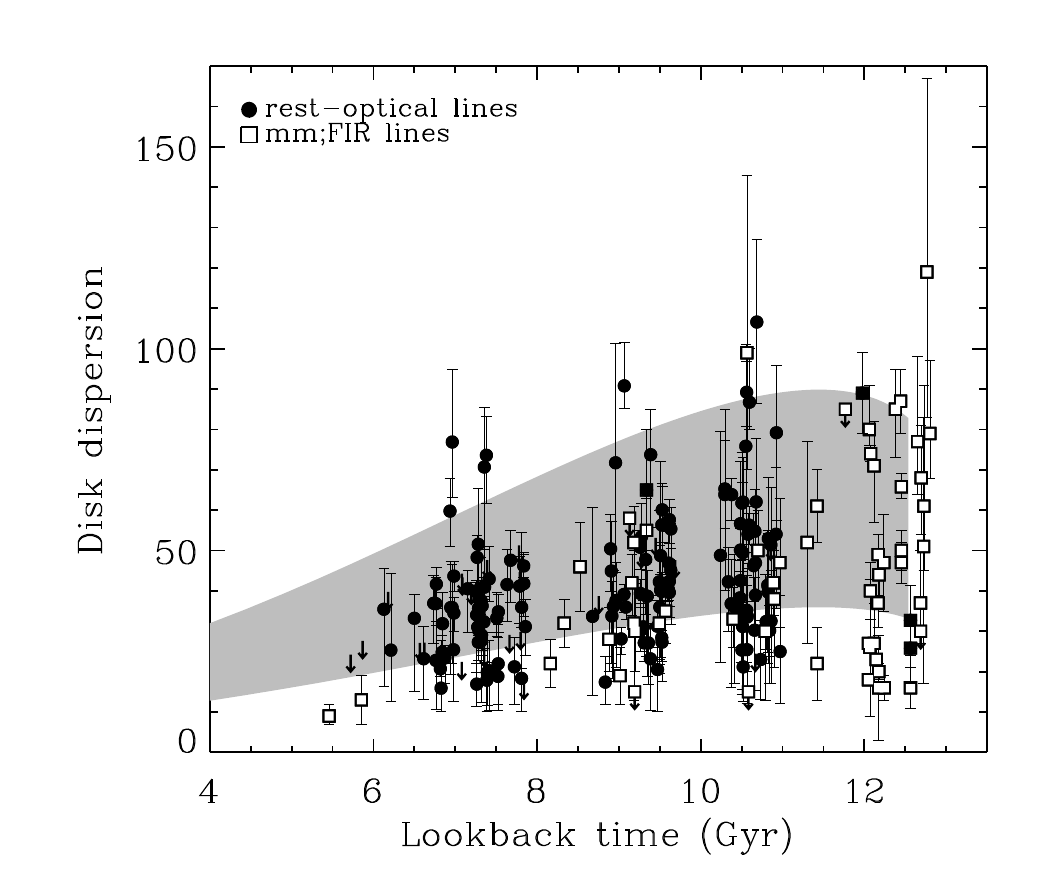}
\caption{Same as Fig.\ref{fig.Tkeyplot} with a linear dispersion axis and a time axis represented in Gyrs.}
\label{fig.age}
\end{figure}

\begin{figure}
\includegraphics[ scale=0.67,trim=0.0cm 0cm 0cm 0cm, clip]{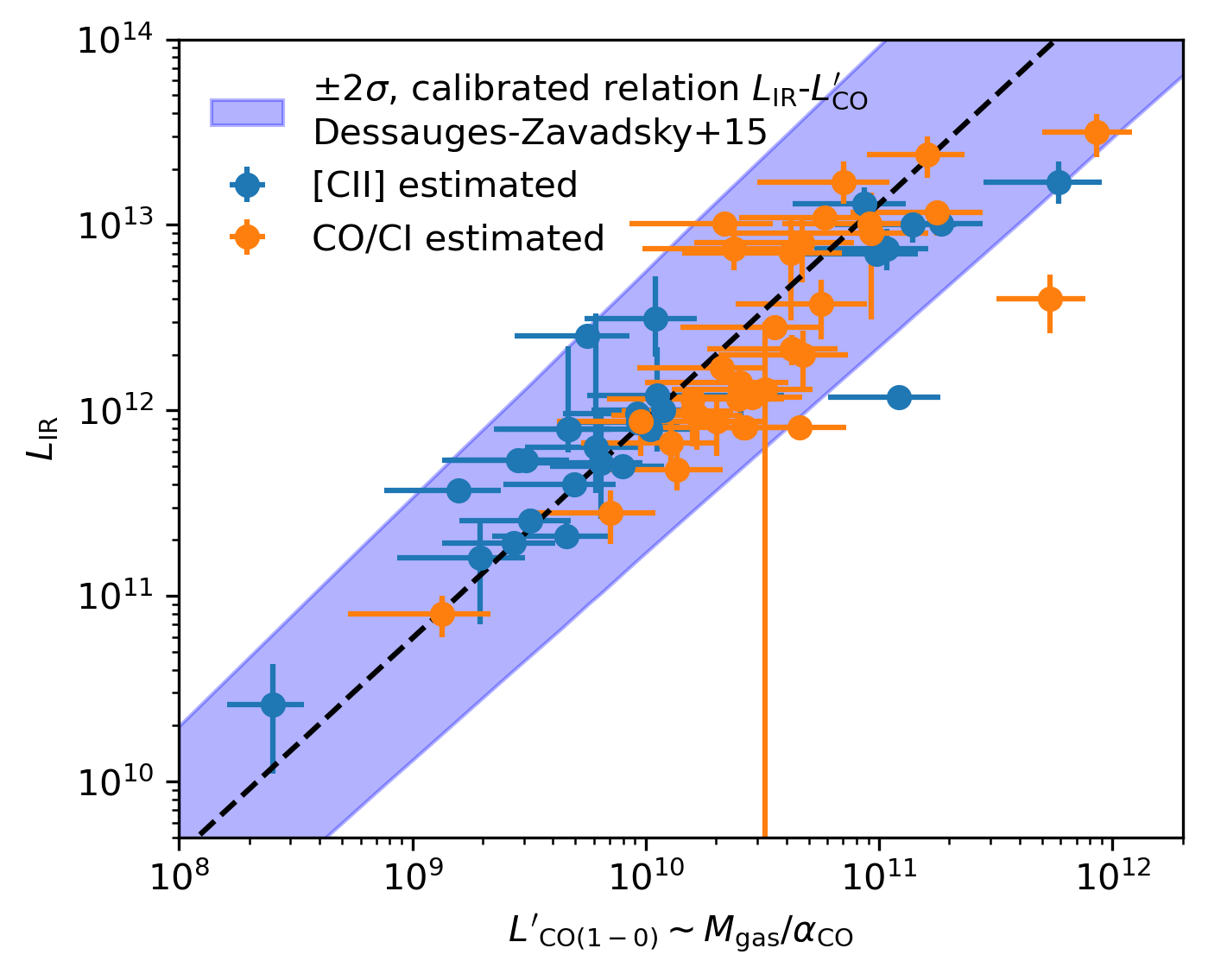}
\caption{Values of $L$(IR) as a function of $L$(CO) for sources with different gas mass tracers discussed in Section~\ref{sec.data}. The relation calibrated in \protect\cite{2015A&A...577A..50D} is shown by the black dashed line and corresponding 2$\sigma$ region in purple shading.}
\label{fig.molgas_cal}
\end{figure}

We reproduce Fig.~\ref{fig.Tkeyplot} in Fig.~\ref{fig.age} to show the evolution as a direct function of lookback time. In Fig.~\ref{fig.molgas_cal} we show the conversion from infrared luminosity, $L_\mathrm{IR}$, to \Mgas~as described in Section~\ref{sub.mmobs}. 

\begin{table*}
\caption{Compiled kinematic measurementsc\textit{(Full table in published version and online materials)}}
\begin{tabular}{lrrrrrrrr}
\hline
Object & Kinematic Paper & RA & DEC & redshift & resolved line & $\sigma$ & $V/\sigma$ & Measurement\\
 & & & & & & [km s$^{-1}$] & & technique$^\mathrm{a}$\\
\hline
EGS12007881            & \cite{2013ApJ...768...74T}   &  14:18:03.60 & 52:30:22.20 & 1.160 & CO($3-2$) & $32.0\substack{+ 6.0 \\  -6.0}$ & $ 7.3\substack{+2.0 \\ -2.0}$ & data \\    
EGS13003805           &  \cite{2013ApJ...768...74T}  &   14:19:40.10  & 52:49:39.10 & 1.230 & CO($3-2$) & $46.0\substack{+11.0 \\ -11.0}$ & $ 7.8\substack{+2.2 \\-2.2}$  & data \\     
EGS13011166            &  \cite{2013ApJ...768...74T}   &   14:19:45.00 & 52:52:28.00 & 1.530 &   CO($3-2$) & $55.0 \substack{  8.0 \\ -8.0}$ &   $6.7\substack{+ 1.5 \\-1.5}$  & data \\     
EGS4-24985              & \cite{2018ApJ...854L..24U}  & 04:19:26.66 & 52:51:17.00 & 1.400 & CO($3-2$) & $19.0\substack{+ 7.0 \\  -7.0}$ & $ 15.6\substack{+0.4 \\ -0.4}$ & DYSMAL \\
BRI1335-0417            & \cite{2021Sci...372.1201T}   &13:38:03.42 &-04:32:35.02 &  4.407 &  \cii & $71.0\substack{+14.0 \\ -11.0}$ & $ 2.5\substack{+0.6 \\ -0.4}$ &  data     \\
zC400569			 & \cite{2023AA...672A.106L} & ...	&  ...   & 2.240  & CO($4-3$)  & $<15.0$  & $>16.0$ & 3D BAROLO \\
zC488879			& \cite{2023AA...672A.106L} &   ...    &  ...   & 1.470   & CO($3-2$)  & $<15.0$  & $>22.4$ & 3D BAROLO \\
\hline  
\end{tabular}\\
\begin{tabular}{l}
$^\mathrm{a}$ Technique used to measure kinematic parameters: data = data driven techniques including using the outer regions; DysmalPy \citep{2021ApJ...922..143P}; \\
GALPAK3D \citep{2015AJ....150...92B}; QubeFit \citep{2020ascl.soft05013N}; 3D BAROLO \citep{2015MNRAS.451.3021D}\\
\end{tabular}\\
\label{tab.kin}
\end{table*}%

\begin{table*}
\caption{Compiled galaxy properties \textit{(Full table in published version and online materials)}}
\begin{center}
\begin{tabular}{lrrrrrrr}
\hline
Object & $\log(M_*/$ & Source & $\log(M_\mathrm{gas}/$ & $M_\mathrm{gas}$ & Source & SFR & Source \\
           & M$_{\odot}$])&  &  M$_{\odot}$]) & tracer &  & [\sfrunits] &  \\
\hline
EGS12007881         & 10.72 & \cite{2013ApJ...768...74T} &  10.86 & CO($3-2$) & \cite{2013ApJ...768...74T} & 94.0 &\cite{2013ApJ...768...74T}\\
EGS13003805         & 11.23 & \cite{2013ApJ...768...74T} &  11.31 & CO($3-2$) &\cite{2013ApJ...768...74T} & 200.0 &\cite{2013ApJ...768...74T}\\
EGS13011166          & 11.08 & \cite{2013ApJ...768...74T} &  11.39 & CO($3-2$) & \cite{2013ApJ...768...74T} & 373.0 &\cite{2013ApJ...768...74T}\\
EGS4-24985	       & 10.87 & \cite{2018ApJ...854L..24U} & 10.84 & CO($3-2$) &   \cite{2018ApJ...854L..24U}  & 98.8 &   \cite{2018ApJ...854L..24U} \\ 
BRI1335-0417       & ...   & ... & 10.75 & CO($2-1$) & \cite{2016ApJ...830...63J} & 1700.0 & \cite{2023MNRAS.523.4654T}  \\ 
zC400569			&  11.30 & \cite{2019ApJS..244...40L}  & 11.07 & CO($3-2$) & \cite{2023AA...672A.106L}  &   81.0 & \cite{2019ApJS..244...40L}  \\
zC488879			&  11.79 & \cite{2019ApJS..244...40L}  & 10.84 & CO($2-1$) & \cite{2023AA...672A.106L}  & 115.0 & \cite{2019ApJS..244...40L} \\
\hline
\end{tabular}
\end{center}
\label{tab.prop}
\end{table*}%

\section{Role of main sequence offset}
\label{app.dMS}
\begin{figure*}
\includegraphics[ scale=0.725, angle=90, trim=8.6cm 0cm 0cm 0.0cm, clip]{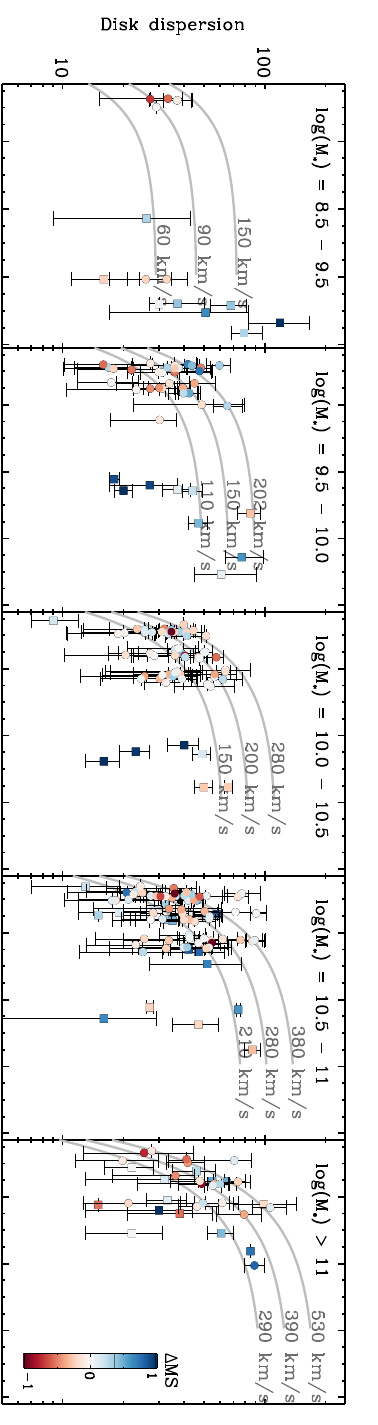}
\includegraphics[ scale=0.725, angle=90, trim=7.6cm 0cm 0.1cm 0.0cm, clip]{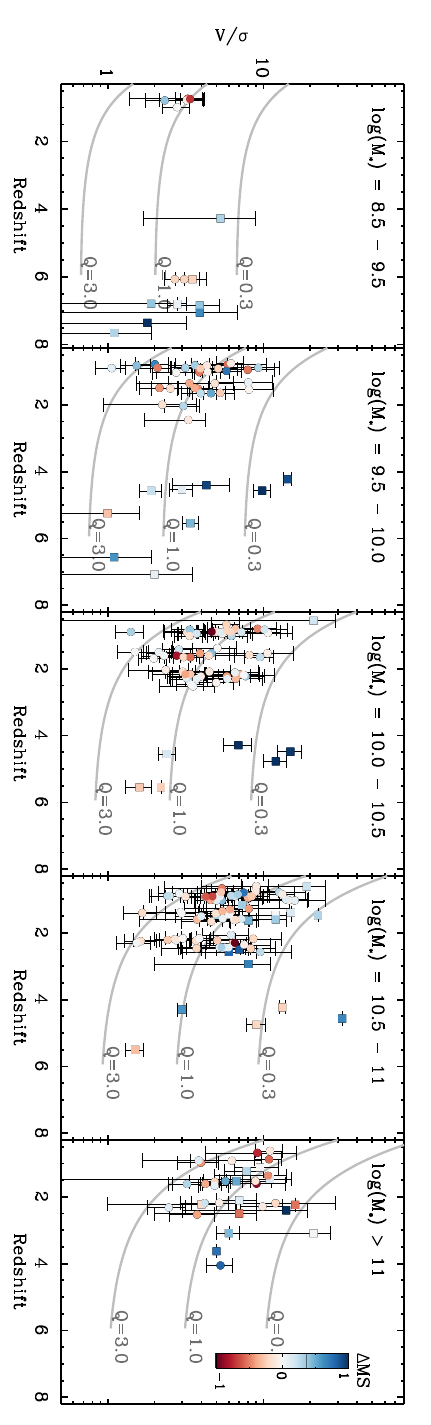}
\caption{Same as Fig.\ref{fig.Tmass} with data points colour-coded by $\Delta$MS.}
\label{fig.Tmass_dMS}
\end{figure*}

\begin{figure*}
\includegraphics[ scale=0.365,trim=1.0cm 0.0cm 0cm 0.0cm, clip]{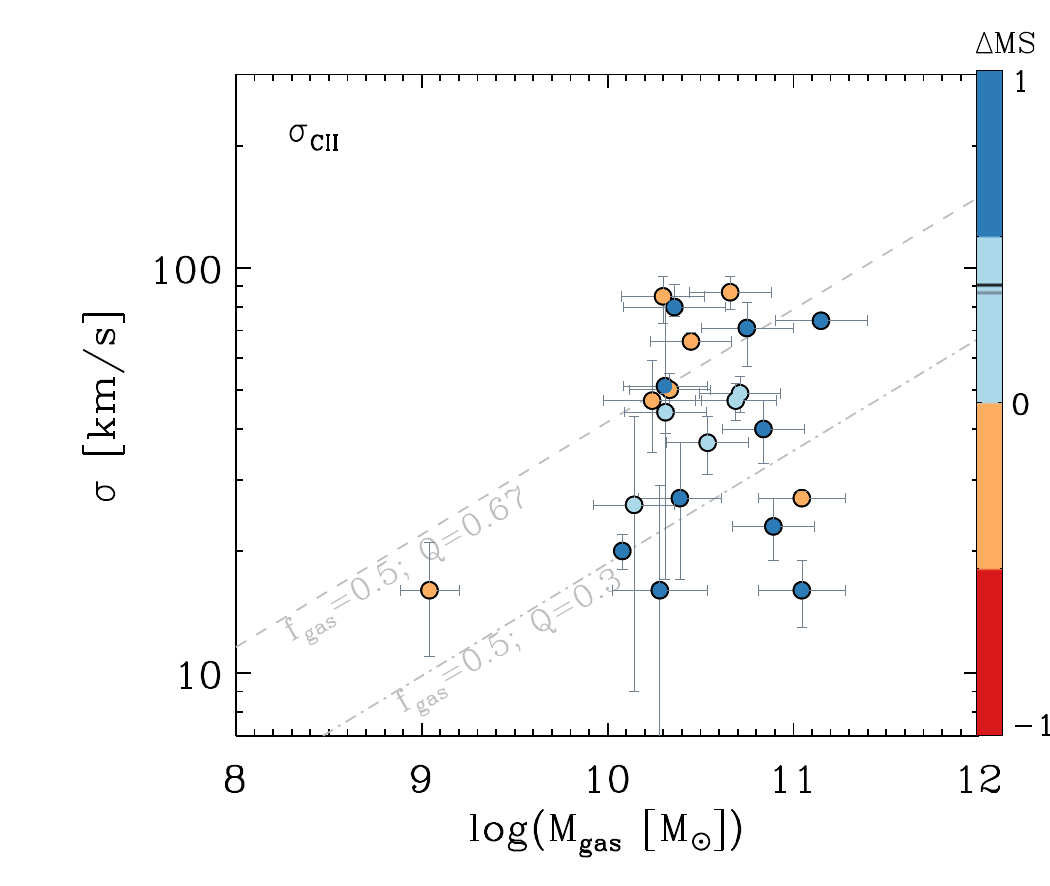}
\includegraphics[ scale=0.365,trim=2.2cm 0.0cm 0cm 0.0cm, clip]{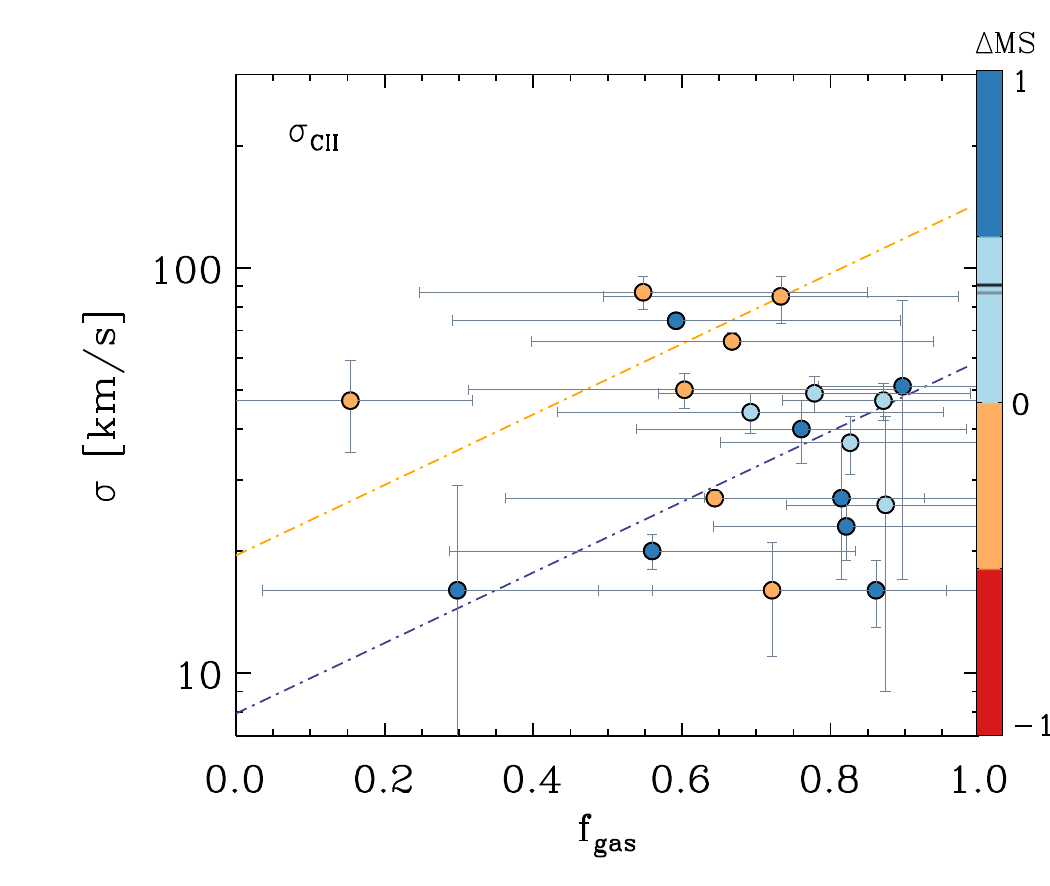}
\includegraphics[ scale=0.365,trim=2.2cm 0.0cm 0cm 0.0cm, clip]{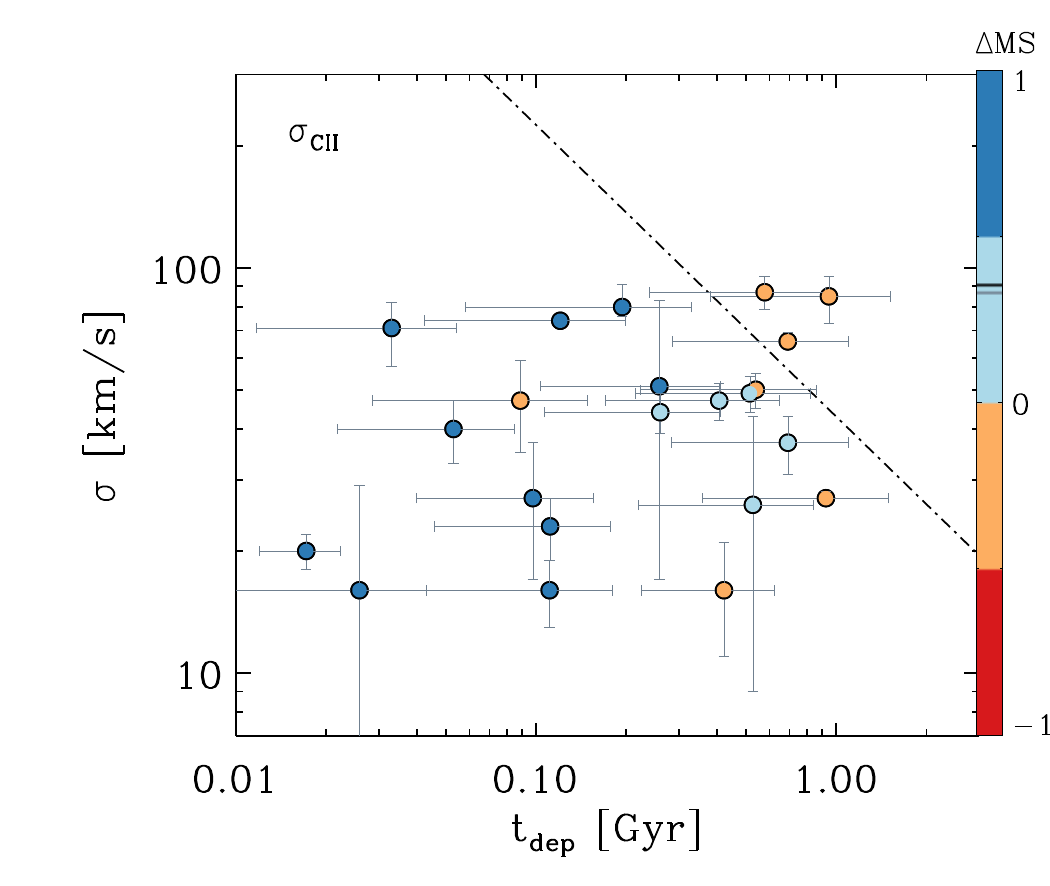}
\caption{Same as Fig.\ref{fig.gas2} with data points colour-coded by $\Delta$MS.}
\label{fig.gas2_dMS}
\end{figure*}

There are many competing factors when exploring the shape and scatter of $\sigma(z)$. For simpicity, the main text focuses on MS galaxies, however galaxies offset from the MS may not be well represented by the model due to the connection with molecular gas content. Here we reproduce Fig.~\ref{fig.Tmass} and Fig.~\ref{fig.gas2} with data points colour-coded by offset from the MS. A clear trend with $\Delta$MS is not obvious from Fig.~\ref{fig.Tmass_dMS} likely due to the number of variables that can cause scatter in the kinematic measurements. In the lower panels, the outliers with high $V/\sigma$ are offset above the MS, contrary to expectations from Section~\ref{sec.toomre}. This may result from the different phases traced by \cii in different galaxies (Section~\ref{sub.phase}). In Fig.~\ref{fig.gas2_dMS}, a reproduction of Fig.~\ref{fig.gas2}, the trends seen in the main text hold. Galaxies with high SFRs tend to have large $\Delta$MS values producing a similar rough separation. 

\bsp	
\label{lastpage}
\end{document}